\documentclass[twocolumn,trackchanges]{aastex701}

\usepackage{graphicx}	
\usepackage{amsmath}	
\usepackage{amssymb}	
\usepackage{xcolor}
\usepackage{color}
\usepackage{soul}
\usepackage{float}
\usepackage{tabularx}
\usepackage{placeins}
\usepackage{afterpage}

\newcommand{\xmm}{XMM-Newton}  

\begin{document}

\title[X-ray reprocessing in ULXs]{Reprocessing of X-ray emission in Ultra-Luminous X-ray sources}

\author{Manish Kumar}
\affiliation{Raman Research Institute, C.V. Raman Avenue, Bangalore 560080, Karnataka, India}
\email[show]{manishk@rrimail.rri.res.in}

\author[orcid=0000-0003-0366-047X]{Rahul Sharma}
\email{rahul.sharma@iucaa.in}
\affiliation{Inter-University Centre for Astronomy and Astrophysics (IUCAA), Ganeshkhind, Pune 411007, India}
\affiliation{Raman Research Institute, C.V. Raman Avenue, Bangalore 560080, Karnataka, India}

\author{Biswajit Paul}
\affiliation{Raman Research Institute, C.V. Raman Avenue, Bangalore 560080, Karnataka, India}
\email{bpaul@rri.res.in}

\begin{abstract}
Most of the Ultraluminous X-ray Sources (ULXs) harbor either a neutron star or a stellar-mass black hole as a compact object accreting at super-Eddington rates. 
In spite of having such a high accretion rate, the reprocessed emission in ULXs is quite meagre compared to that observed in Galactic X-ray binaries, except for some absorption lines in the winds. In this work, we use archival Chandra and \xmm\ observations to investigate the extent of reprocessed emission in ULXs through three diagnostics: (i) searches for Fe $\rm K\alpha$ lines in bright well-known ULXs and Ultraluminous X-ray Pulsars (ULXPs), (ii) evolution of hardness ratio around the eclipse transitions in the eclipsing ULXs, and (iii) the flux ratio between eclipse and out-of-eclipse (OOE) phases in eclipsing ULXs.
We constrain the equivalent width (EW) of the Fe K$\alpha$ line, with upper limits of $\sim$11–20 eV for a fixed line centroid (6.4 keV) and width (0.1 keV). Allowing the centroid and width to vary yields more conservative upper limits of up to $\sim$50 eV.
Furthermore, we have not observed any significant changes in the hardness ratio in eclipsing ULXs during ingress or egress, whereas in Galactic eclipsing X-ray binaries the hardness ratio increases during the transitions. Finally, the reprocessing efficiency (eclipse-to-OOE flux ratio) is found to be larger in ULXs compared to Galactic eclipsing X-ray binaries. Based on these results, we discuss the possibility of a metal-poor or highly ionized environment surrounding the ULXs, which suppresses reprocessed emission features.\\

\end{abstract}




\section{Introduction}
X-ray sources with extremely high luminosities, positioned off-center in nearby galaxies and exceeding the Eddington luminosity for stellar-mass compact objects, were first observed by the Einstein Observatory \citep{Fabbiano1989}. 
Primarily identified as off-nuclear sources and hence distinct from active galactic nuclei (AGN), follow-up observations with \emph{ROSAT} \citep{ROSAT_Truemper_1982}, \xmm\ \citep{Jansen2001}, and \emph{Chandra} \citep{chandra_Weisskopf_2002} X-ray observatories later confirmed this nature.
This pivotal finding led to the classification of a new category of X-ray sources, now referred to as ultraluminous X-ray sources (ULXs) \citep{Makishima_2000, Fabrika, King2023}. So far, approximately 1800 ULXs have been identified \citep{Earnshaw_ULX_catalogue, walton_2022, ULX_catalogue_Tranin_2024}.

The luminosity of the persistent and transient ULXs in their bright state is in the range of $10^{39} - 10^{41}$ erg s$^{-1}$ in the X-ray energy band, which is much greater than the Eddington luminosity for stellar-mass compact objects. In some of the ULXs, like M82 X-2, NGC 5907 ULX-1, NGC 7793 P13, NGC 300 ULX-1, M51 ULX-7, NGC 1313 X-2, coherent pulsations have been detected, identifying them as ultraluminous X-ray Pulsars (ULXPs) \citep{Bachheti2014, Furst_7793, Israel_5907, Israel_7793, Carpano, sathyapraksh_1313x2, Rodríguez_Castillo_2020}. The most common interpretation is that many of the ULXs are stellar-mass compact objects, either black holes (BHs) or neutron stars (NSs), accreting at super-Eddington rates. In some ULXPs, orbital parameters have been measured, allowing estimation of the mass function and, consequently, the minimum companion mass. For instance, in M82 X-2, \cite{Bachheti2014} calculated the mass function $f=2.1 \, M_{\rm \odot}$ and hence the minimum mass for the companion star is $\rm M_{c} \, > \, 5.2 \,M_{\rm \odot}$ suggesting a high-mass X-ray binary (HMXB). Similarly, M51 ULX-7 is also inferred to have a high-mass companion with a minimum mass $\rm M_{c} \, > \, 8 \, M_{\rm \odot} $ \citep{Rodríguez_Castillo_2020}. \cite{5907_ulx1_orbit_belfiore} suggested that NGC 5907 ULX-1 is a HMXB system viewed nearly face-on, with an orbital inclination $i < 5^{\rm^{o}}$. Furthermore, the presence of long-duration eclipses in some ULXs suggests that high-mass companions may be a common feature among many, if not all, ULXs \citep{Urquhart2016}. The high X-ray luminosity and presence of a high-mass companion indicate ULXs to have a large amount of material surrounding the X-ray source, and ULXs are therefore expected to show strong signatures of X-ray reprocessing in the surroundings of the compact object.


One common approach to study the environment surrounding a compact object is to examine the Fe K$\alpha$ line, which arises from the fluorescence of iron atoms when illuminated by X-rays from the central source. This emission is generally associated with reflection from surrounding material, such as the accretion disk, stellar wind, or other structures in the vicinity of the compact object.
The Fe K$\alpha$ line has been extensively studied in Galactic X-ray binaries, where its equivalent width (EW) spans a wide range from a few eV to $\approx$1.7 keV \citep{Torrejon_2010_iron_line, C_Ng_LMXB, garcia_xmm_newton_Fe_k_alpha_HMXB, Pradhan_SFXT_SGHMXB}. The same material, responsible for producing the iron line, can also contribute to absorption along the line of sight.
Galactic classical supergiant high-mass X-ray binaries (SG HMXBs) exhibit a clear correlation between the column density and the EW of the iron line. This is expected, as the primary accretion mechanism in these systems is via the stellar wind from the companion star. At larger column density, more material is exposed to X-ray illumination, leading to a stronger iron line and a higher EW \citep{Torrejon_2010_iron_line, garcia_xmm_newton_Fe_k_alpha_HMXB, Pradhan_SFXT_SGHMXB}. However, the average EW of the iron line in supergiant fast X-ray transients (SFXTs) is generally lower than that observed in classical SG HMXBs \citep{Pradhan_SFXT_SGHMXB}. \cite{C_Ng_LMXB} found the EW of the iron line for low-mass X-ray binaries (LMXBs) ranging between 17 eV and 189 eV, which is, on average, smaller than that observed in HMXBs. 


Another way to probe X-ray reprocessing in X-ray binaries is through eclipse and eclipse-transition phases, where the direct emission from the compact object is largely obscured by the companion star and the observed spectrum is dominated by scattered and fluorescence emission.
These reprocessing effects manifest observationally through characteristic spectral and flux variations.
Galactic eclipsing HMXBs show an increase in hardness ratio during ingress/egress \citep{Vela_X1_HMXB_Abalo_2024, Ajith_Cen_X3, Tamang_Vela_X1}, which indicates a larger absorption column density, hence more absorption of soft photons near the compact object, during ingress/egress. In addition, the X-ray flux in Galactic HMXBs significantly drops during the eclipse phase. \cite{Nafisa_HMXB_eclipsing, Aftab_LMXB} found that the ratio of flux for out-of-eclipse (OOE) to eclipse varies between $\sim$8--237 in HMXBs and $\sim$6--54 in LMXBs, varying from source to source.

In this context, different observables probe distinct physical aspects of X-ray reprocessing. The Fe K$\alpha$ emission line probes the environment illuminated by its X-ray emission from the compact object; the evolution of the hardness ratio during eclipse transitions primarily traces spectral variations caused by changes in absorption and/or scattering in the surrounding medium; and the flux ratio between eclipse and out-of-eclipse phases further provides a measure of the relative contribution of reprocessed emission compared to the intrinsic source flux. 

In this work, we aim to study the above three complementary aspects of X-ray reprocessing in ULXs: the Fe $\rm K\alpha$ emission line, the hardness ratio around the eclipse transitions, and the flux ratio during the eclipse to OOE phase.
To study the iron line, we selected a sample of the brightest ULXs (see Appendix \ref{apx:selection} for the sample selection procedure), comprising seven well-known ULXs and five known ULXPs.
To study the hardness ratio and flux ratio, we selected all five known eclipsing ULXs: CG X-1, CXOM51 J132943.3+471135 (M51 S1), CXOM51 J132946.1+471042 (M51 S2), CXOM51 J132940.0+471237 (M51 ULX-1), and CXOM51 J132939.5+471244 (M51 ULX-2). Archival data from \xmm\ and Chandra observatories are used in this study.

\section{Observations \& Source Selection}

We investigate the iron line aspect of the well-known bright ULXs with long-exposure observations with XMM-Newton. We utilized all available XMM-Newton observations for the selected 7 ULXs and 5 ULXPs. A similar study was done previously for sources Ho IX X-1 and NGC 1313 X-1 by \cite{NGC_1313_X1_Ho_IX_X1_walton}, which reported the upper limit on the EW of iron line of $\sim$30 eV and $\sim$50 eV, respectively. The summary of exposure time is mentioned in Table \ref{tab:total_GTE}, and the detailed observation log in Table \ref{tab:observation_log}. The source NGC 300 ULX-1 was also observed in 2010 and showed a prominent iron line feature near 6.4 keV \citep{Carpano}. The spectrum obtained from this particular observation is remarkably different from that of other ULXs, and no pulsations were detected. Therefore, this dataset is excluded from the present analysis.

\begin{table}
\caption{Total good-time exposure (GTE) for each ULX and ULXP included in the search for iron emission line, based on all available XMM-Newton observations used in the analysis.}
\label{tab:total_GTE}
\fontsize{10}{10}\selectfont
\resizebox{\columnwidth}{!}{
\renewcommand{\arraystretch}{1.5}
\hskip-2.0cm\begin{tabular}{ccccc}
\hline
Source Name & RA & Dec & GTE (ks) & $N_{\rm H}$ ($\rm 10^{22} \, cm^{-2}$) \\
\hline
M51 ULX-7$^{\rm a1}$ &  $\rm 13^{h}30^{m}1.02^{s}$ &   $\rm 47^{\circ}13^{'}43.80^{''}$ & 493.53 & $\rm 0.099^{b1}$  \\
NGC 300 ULX-1$^{\rm a2}$ &  $\rm 00^{h}55^{m}4.85^{s}$ &   $\rm -37^{\circ}41^{'}43.50^{''}$ & 141.07 & $\rm 0.066^{b1}$ \\
NGC 1313 X-2$^{\rm a3}$ &  $\rm 3^{h}18^{m}22.00^{s}$ &   $\rm -66^{\circ}36^{'}4.30^{''}$ & 883.13 & $\rm 0.218^{b1}$ \\
NGC 5907 ULX-1$^{\rm a4}$ &  $\rm 15^{h}15^{m}58.62^{s}$ &   $\rm 56^{\circ}18^{'}10.30^{''}$ & 779.24 & $\rm 0.725^{b1}$ \\
NGC 7793 P13$^{\rm a5}$ &  $\rm 23^{h}57^{m}50.90^{s}$ &   $\rm -32^{\circ}37^{'}26.60^{''}$ & 412.84 & $\rm 0.091^{b1}$  \\
Circ ULX-5$^{\rm a6}$ &  $\rm 14^{h}12^{m}39.00^{s}$ &   $\rm -65^{\circ}23^{'}34.00^{''}$ & 301.67 & $\rm 0.610^{b2}$ \\
Holmberg II X-1$^{\rm a7}$ &  $\rm 08^{h}19^{m}28.99^{s}$ &  $\rm 70^{\circ}42^{'}19.37^{''}$ & 232.64 & $\rm 0.059^{b3}$ \\
Holmberg IX X-1$^{\rm a8}$ &  $\rm 09^{h}57^{m}53.20^{s}$ &  $\rm 69^{\circ}03^{'}48.30^{''}$ & 156.46 & $\rm 0.119^{b3}$ \\
NGC 1313 X-1$^{\rm a9}$ &  $\rm 03^{h}18^{m}20.00^{s}$ &  $\rm -66^{\circ}29^{'}11.00^{''}$ & 884.82 & $\rm 0.253^{b4}$ \\
M33 X-8$^{\rm a10}$ &  $\rm 01^{h}33^{m}50.60^{s}$ &  $\rm 30^{\circ}39^{'}31.00^{''}$ & 264.53 & $\rm 0.187^{b5}$ \\
NGC 4559 ULX-7$^{\rm a11}$ &  $\rm 12^{h}35^{m}51.71^{s}$ & $\rm 27^{\circ}56^{'}4.10^{''}$ & 164.60 & $\rm 0.150^{b6}$ \\
IC 342 X-2$^{\rm a12}$ & $\rm 03^{h}46^{m}15.70^{s}$ & $\rm 68^{\circ}11^{'}12.00^{''}$ & 85.39 & $\rm 1.84^{b7}$ \\
\hline
\multicolumn{5}{p{0.7\textwidth}}{1. References for the source co-ordinates and previous detailed studies: $^{\rm a1}$\citep{Rodríguez_Castillo_2020, Brightman_2022}, 
$^{\rm a2}$\citep{Carpano}, 
$^{\rm a3}$\citep{Robba}, 
$^{\rm a4}$\citep{Furst_5907, Israel_5907}, 
$^{\rm a5}$\citep{Furst_7793, Israel_7793}, 
$^{\rm a6}$\citep{Circ_ULX5_Mondal}, 
$^{\rm a7}$\citep{Ho_II_X1_Barra}, 
$^{\rm a8}$\citep{NGC_1313_X1_Ho_IX_X1_walton, Ho_IX_X1_Walton_2017}, 
$^{\rm a9}$\citep{NGC_1313_X1_Ho_IX_X1_walton}, 
$^{\rm a10}$\citep{M33_X8_West_2018}, 
$^{\rm a11}$\citep{NGC_4559_X7_rare_flaring_Pintore}
$^{\rm a12}$\citep{IC_342_X2_Rana_2015}}
\\ 
\multicolumn{5}{p{0.7\textwidth}}{2. Column density ($\rm N_{H}$) values for the curve of growth are adopted from the previous literature as follows: $^{\rm b1}$\citet{Manish_ULXPs}, 
$^{\rm b2}$\citet{Circ_ULX5_Mondal}, 
$^{\rm b3}$\citet{Pintore_ULX_2014}, 
$^{\rm b4}$\citet{NGC1313_X1_Walton_2020}, 
$^{\rm b5}$\citet{M33_X8_West_2018}, 
$^{\rm b6}$\citet{NGC_4559_X7_rare_flaring_Pintore}
$^{\rm b7}$\citet{IC_342_X2_Rana_2015}
}\\

\end{tabular}}
\end{table}

We selected five known eclipsing ULXs (CG X-1, M51 S1, M51 S2, M51 ULX-1, and M51 ULX-2) to study the hardness ratio with orbital phase, mainly during the ingress/egress of the eclipses and the flux ratio between the eclipse and OOE phases. Since M51 S1 and M51 S2 are well resolved with \xmm\ instrument, we utilized the \xmm\ observations for these two sources. For CG X-1, M51 ULX-1, and M51 ULX-2, we used Chandra observations, as listed in Table \ref{tab:eclipsing_obs_log}.

We note that only a subset of the available \emph{Chandra} and \emph{XMM-Newton} observations is listed in Table \ref{tab:eclipsing_obs_log}. These observations were selected based on their coverage of eclipse or partial eclipse phases, together with data quality considerations. For CG X-1, we utilized a \emph{Chandra} observation which uniquely covers $\sim$6 eclipse cycles, 
while the remaining observations either cover only partial eclipses or are not sufficiently bright for detailed analysis. For M51 S1 and S2, all available \emph{XMM-Newton} observations exhibiting eclipse or partial eclipse features were included. 
For M51 ULX-1, both observations 13813 and 13814 exhibit partial eclipses, while for M51 ULX-2, a partial eclipse is observed only in observation 13813. However, the data during the eclipse phases are limited by low counting statistics for both sources. Therefore, we retained observation 13813 as a representative dataset for the spectral analysis of both sources, while observation 13814 for M51 ULX-1 was excluded as it did not yield results significantly different from those obtained using observation 13813.

\begin{table}
  \caption{Observation log for the eclipsing ULXs analyzed in this work, including the datasets used for studying hardness ratio evolution and flux variations between eclipse and OOE phases.}
 \label{tab:eclipsing_obs_log}
\fontsize{10}{10}\selectfont
\resizebox{\columnwidth}{!}{
\renewcommand{\arraystretch}{1.5}
\hskip-3.0cm\begin{tabular}{cccccccc}
\hline
\hline
Source & Co-ordinates & Instrument & Obs Id & MJD & Source & Total \\
Name & &  &  & & Region & GTE (ks) \\
\hline
\hline
CG X-1$^1$ & RA = $\rm 14^{h}13^{m}12.21^{s}$ & Chandra & 12823 & 55547.75 & $4^{''}$ & 142.87 \\
  & Dec = $\rm -65^{\circ}20^{'}13.7^{''}$ & & & & \\ 
\hline
M51 S1$^2$ & RA = $\rm 13^{h}29^{m}43.32^{s}$ & XMM-Newton & 0824450901 & 58251.92 & $20^{''}$ & 64.73 \\
  & Dec = $\rm +47^{\circ}11^{'}34.9^{''}$  & XMM-Newton & 0852030101 & 58675.49 & $20^{''}$ & 58.78 \\
\hline
M51 S2$^2$ & RA = $\rm 13^{h}29^{m}46.13^{s}$  & XMM-Newton & 0830191501 & 58282.10 & $15^{''}$ & 51.59 \\
  & Dec = $\rm +47^{\circ}10^{'}42.3^{''}$  & XMM-Newton & 0830191601 & 58284.09 & $15^{''}$ & 51.66 \\
  &   & XMM-Newton & 0852030101 & 58675.49 & $20^{''}$ & 58.78 \\
\hline
M51 ULX-1$^3$ & RA = $\rm 13^{h}29^{m}39.94^{s}$  & Chandra & 13813 & 56179.74 & $4^{''}$ & 166.07 \\
  & Dec =  $\rm +47^{\circ}12^{'}36.6^{''}$  &  & &  &  \\
\hline
M51 ULX-2$^3$ & RA = $\rm 13^{h}29^{m}39.44^{s}$  & Chandra & 13813 & 56179.74 & $4^{''}$ & 166.07  \\
  & Dec =  $\rm +47^{\circ}12^{'}43.3^{''}$  &  & &  &  \\
\hline

\multicolumn{7}{l}{$^1$\citet{Cg_X1_Qiu_2019}, $^2$\citet{M51_S1_S2_Wang}, $^3$\citet{Urquhart2016}.}
\end{tabular}}
\end{table}

\section{Data Reduction}
\subsection{XMM-Newton}

\xmm\ is a space-based X-ray observatory operated by the European Space Agency (ESA) \citep{Struder2001, Turner2001}. For the present analysis, we use only EPIC-pn data 
which provides a higher effective area and 
counting statistics compared to MOS, making it more suitable for detecting faint spectral features. Inclusion of MOS data does not significantly improve the constraints and may introduce additional cross-calibration uncertainties. 

For the reduction of \xmm\ data, we used the \xmm\ Science Analysis System (SAS) version 20.0.0 and followed the SAS Data analysis threads. First, the raw event files were extracted using \texttt{epproc} for the listed observations. Then we performed the background flaring correction, if required, depending on the observations. The clean event files are barycenter corrected using the SAS task \texttt{barycen}.
Source events were extracted from circular regions centered on the source coordinates, with radii as specified in Tables \ref{tab:observation_log} and \ref{tab:eclipsing_obs_log}. The background events were extracted from a circular region with a radius 1.5 times that of the source region (2 times for M51 S1 and S2), selected in close proximity to the source position on the same CCD and free from contamination by other sources.
Instrument response files (rmf) are generated using the tasks \texttt{rmfgen}\footnote{\url{https://www.cosmos.esa.int/web/xmm-newton/sas-thread-epatplot}}, which also include the pile-up corrections (if any), and ancillary response files (arf) are generated using the tasks \texttt{arfgen}. We use the tool \texttt{epiccombine} and \texttt{addspec} for combining the spectra. Spectra are grouped using \texttt{ftgrouppha} with signal-to-noise based binning (\texttt{grouptype=snmin}), adopting \texttt{groupscale=4.0} (S/N $\geq 4$ per bin) for the brightest ULXs, while for eclipsing ULXs the grouping is adjusted within \texttt{groupscale=3-5} to ensure statistically meaningful bins without compromising spectral resolution. We extracted the source and background light curves for all good single and double events ($\rm PATTERN<=4 $) using \texttt{evselect} and obtained final background-subtracted light curves using \texttt{epiclccorr}. 
In some observations, the source falls partially on a CCD chip gap. However, this does not significantly affect the spectrum around 6.4 keV, except for some loss of flux, which is accounted for through ARF generated using the task \texttt{arfgen} \citep{XMM_SAS2025}. We therefore retained these observations in our Fe K$\alpha$ line analysis.

\subsection{Chandra}
The Chandra X-ray Observatory, a flagship mission of NASA, is dedicated to explore the high-energy universe through X-ray observations with a very high angular resolution \citep{chandra_Weisskopf_2002, chandra_acis_garmire}. We use data from the \emph{Chandra} X-ray Observatory, obtained in the ACIS-S configuration.

For event file extraction, we used the Chandra Interactive Analysis of Observations (\texttt{CIAO}) software package version $4.15$ together with the corresponding calibration database \texttt{CALDB} version 4.10.7 \citep{ciao_chandra_Fruscione}. We reprocessed the data using the \texttt{CIAO} task \texttt{chandra\_repro}. We performed the flaring and barycenter correction using task \texttt{deflare} and \texttt{axbary}, respectively. Source light curves were extracted with \texttt{dmextract} from circular regions of radius $4^{\prime\prime}$ centered on the source coordinates listed in Table \ref{tab:eclipsing_obs_log}.
Background regions were selected as circle with radii of $15^{\prime\prime}$ for CG X-1 and $10^{\prime\prime}$ for M51 ULX-1/ULX-2, and were chosen near the source positions on the same CCD, avoiding contamination from other sources.
Energy spectra were extracted from the reprocessed event files using the \texttt{specextract} task. Spectra are grouped using \texttt{ftgrouppha} with signal-to-noise based binning, with \texttt{groupscale} adjusted in the range 3–5 depending on the data quality.

\section{Analysis and Results}

\subsection{Search for iron emission line}
\label{search_Fe_line}

To examine the possible presence of an iron line feature in the combined EPIC-pn spectra of individual sources, we restrict the analysis to the energy range
from 5.0 keV to 8.0 keV, which encompasses the Fe K$\alpha$ band (6.4--7 keV) while minimizing contamination from softer spectral components. Spectral fitting was performed using the \textsc{xspec} software package \citep{Arnaud96}. Previous studies suggest that the spectra of most ULXs are likely dominated by a power-law in the energy range considered above \citep{walton_2022, King2023}. 
Accordingly, we model the continuum in this limited energy band using a power law. We emphasize that this choice of continuum is adopted to model the local continuum around 6.4 keV, rather than to represent the full physical spectral properties of ULXs.

\begin{figure*}
	\includegraphics[height=15cm, width=18cm]{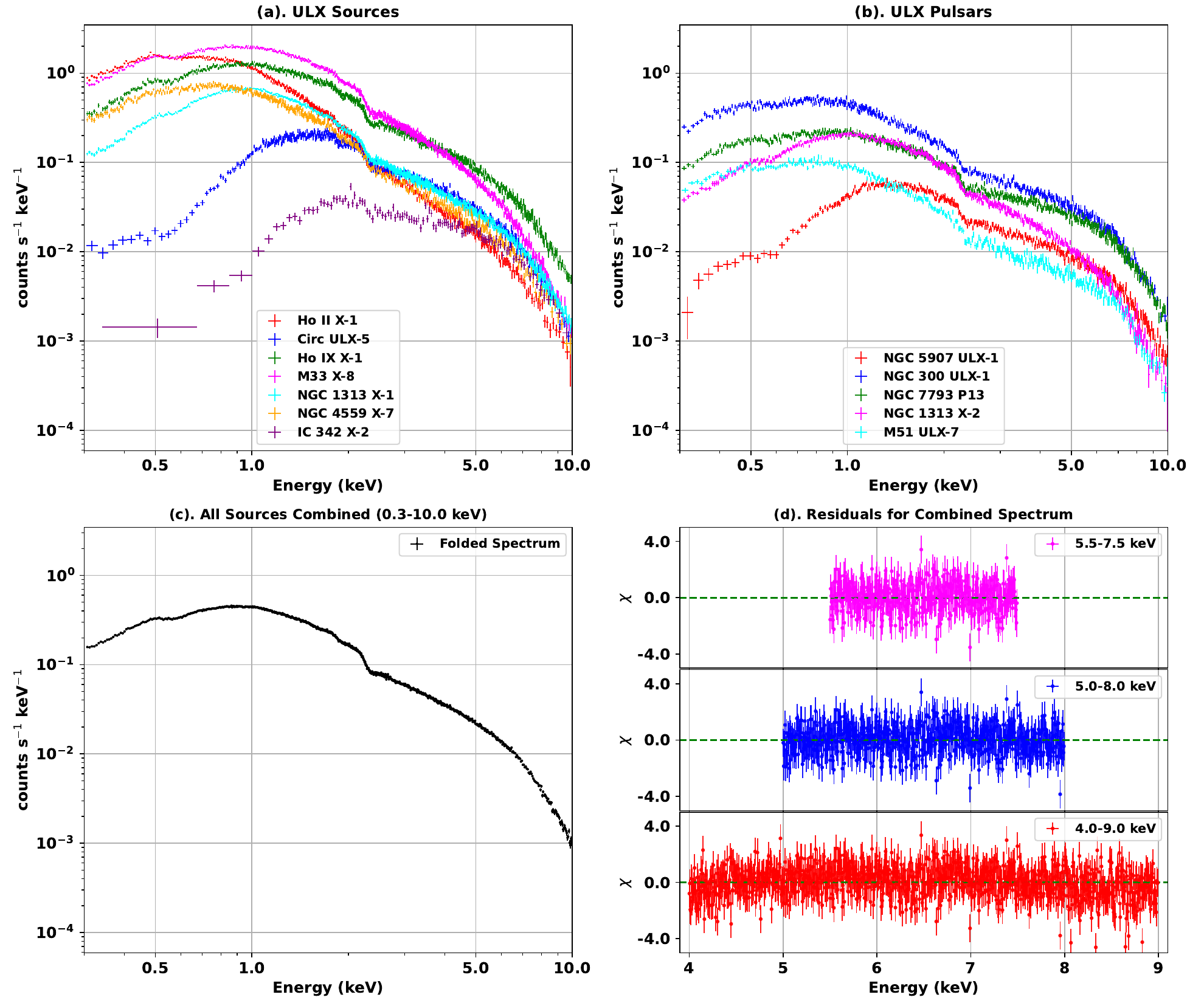}
        \caption{Top panel presents the combined spectra from \xmm\ observations of individual ULX (left) and ULXP (right) sources. The bottom panel (on the left) presents the combined spectra from all observations of 6 ULXs and 5 ULXPs analyzed in this work. On the right, residuals from the best-fit model are shown for the combined spectra in the energy range of 5.5--7.5 keV, 5--8 keV, and 4--9 keV. The spectra are shown with coarser binning for visual clarity, while the residuals reflect the actual binning used in the analysis.}
    \label{all_sources_added_spectrum}
\end{figure*}

To investigate the presence of an iron emission line, we added a Gaussian component to the continuum model, resulting in the final \textsc{xspec} model \texttt{(powerlaw+gauss)}. The top panels of Fig. \ref{all_sources_added_spectrum} show the combined spectra from \xmm\ observations of individual ULXs (left) and ULXPs (right). In all cases, the spectra in the 5--8 keV range are well-described with this simple model, and the best-fit spectral parameters with $\chi^2$/dof values for each source are summarized in Table \ref{tab:fit_en_5_8}. The powerlaw photon index is found to lie between 1.5 and 3.1 for these 12 sources.
Since no significant detection of the line was found, we estimated the 90\% confidence upper limits on the EW by fixing the Gaussian centroid at 6.4 keV and the line width at 0.1 keV. The resulting upper limits on the EW range from 5.4 eV to 73.7 eV across the individual ULX and ULXP spectra (Table \ref{tab:fit_en_5_8}). 

We also attempted to fit the spectra by slightly adjusting the energy range, both narrowing and widening it, to observe the effect on the EW based on the selected energy range. The narrower range used was 5.5--7.5 keV, while the wider range extended from 4.0--9.0 keV. The results, summarized in Table \ref{tab:fit_all_en}, demonstrate that for most sources, the chosen energy range does not significantly influence the upper limit for the EW of the iron line. For all cases, the continuum is fitted using a powerlaw model except for M51 ULX-7 in the energy range 4.0--9.0 keV, in which a powerlaw with high-energy cutoff is used for the continuum fitting. $\rm E_{cut}$  and $\rm E_{fold} $ are frozen to values of 7.2 keV and 3.5 keV, respectively, which are obtained from fitting the full spectrum in the energy range of 0.3--10 keV. Fig. \ref{fig:spec} shows the residuals from the best-fit powerlaw model in the energy range of 5.5--7.5 keV, 5--8 keV, and 4--9 keV for individual ULXs.

Additionally, we fit the combined spectra of all sources, with the best-fit parameters given in Table \ref{tab:fit_en_5_8}. 
The bottom panel of Fig. \ref{all_sources_added_spectrum} shows the combined spectra (left) obtained from all observations of the selected sources, along with the residuals (right) from the best-fit model in the energy range of 5.5--7.5 keV, 5--8 keV, and 4--9 keV. Based on these analyses, we constrain the upper limit on EW of the iron line for the ULXs to be in the range 11--20 eV (Table \ref{tab:fit_all_en}), marking a significant improvement over previous studies.

\subsection{Robustness of the Fe K$\alpha$ Line Constraints}
\label{robustness_Fe_line}
Since the results presented above are based on the combined spectra, it is important to assess whether combining data from different sources/observations, potentially spanning different spectral states, could bias or dilute the detection of a weak Fe K$\alpha$ line. Although stacking improves the counting statistics and hence the sensitivity to faint spectral features, ULXs are known to exhibit spectral variability. To investigate any possible effect of spectral variability on our results, we separated the observations into two spectral classes based on their individual photon indices and independently analysed the combined spectra of these classes. Further, we performed simulations to examine the recovery of weak Fe K$\alpha$ lines from the combined spectra. Finally, we examined the dependence of the derived EW upper limits on the assumed line centroid and width. The results of these analyses are described below.

We first fitted the individual spectra of all 155 observations in the 4.5--8.5 keV energy range using a power-law model. Four observations were excluded because their spectra were too faint to constrain the spectral parameters reliably. 
The excluded observations are M51 ULX7 (ObsID: 0677980701), NGC 5907 ULX1 (ObsID: 0893810301), NGC 7793 P13 (ObsID: 0883780201), and Holmberg II X-1 (ObsID: 0112520901). For the remaining 151 observations, we obtained the photon index from the individual fits and examined its distribution, as shown in Fig.~\ref{fig:all_4_less_greater_than}. Based on the distribution, we divided the observations into two spectral classes: a harder class with $\Gamma < 2.25$ and a softer class with $\Gamma \geq 2.25 $. We then separately combined the spectra within each class and 
constrained the Fe K$\alpha$ line EW over the same three energy ranges considered above. For the harder class, we obtained EW upper limits of $\sim$17--34 eV, while the softer class yielded corresponding upper limits of $\sim$10--15 eV. No significant Fe K$\alpha$ emission is detected in either spectral class (see Fig.~\ref{fig:all_4_less_greater_than}), demonstrating that the absence of line detection is not driven by combining observations with different spectral slopes.

\begin{figure}
    
    \includegraphics[height=7.5cm, width=\columnwidth]{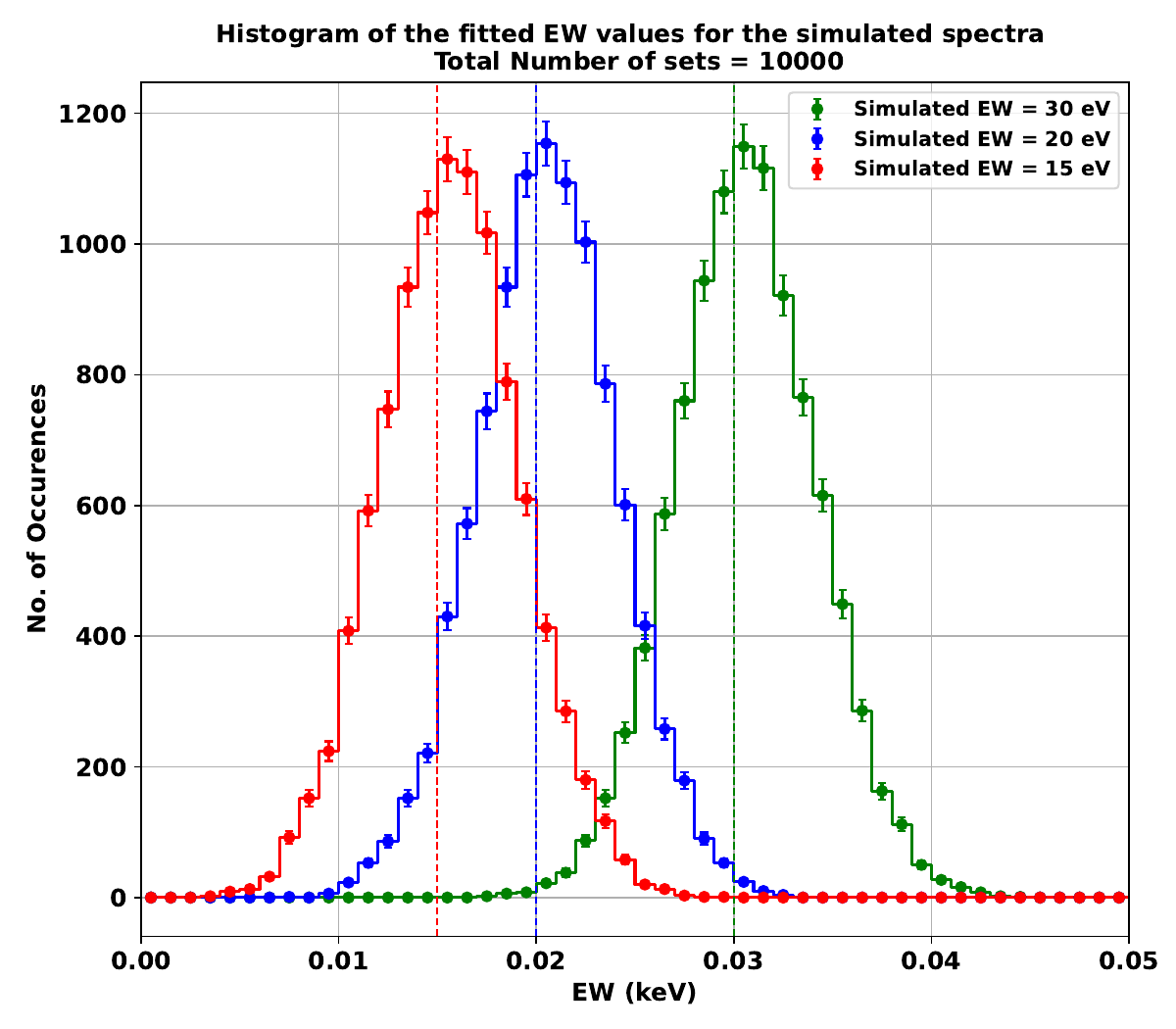}
    \caption{Histogram for the EW recovered from fits to the combined simulated spectra of all sources. Different colors represent simulations in which artificial Fe K$\alpha$ lines with EW values of 15 eV, 20 eV, and 30 eV were inserted into the individual source spectra before combining.}
    \label{fig:simulated_EQW_validation}

\end{figure}

\begin{figure*}
    \centering
    \includegraphics[height=6.8cm, width=0.46\linewidth]{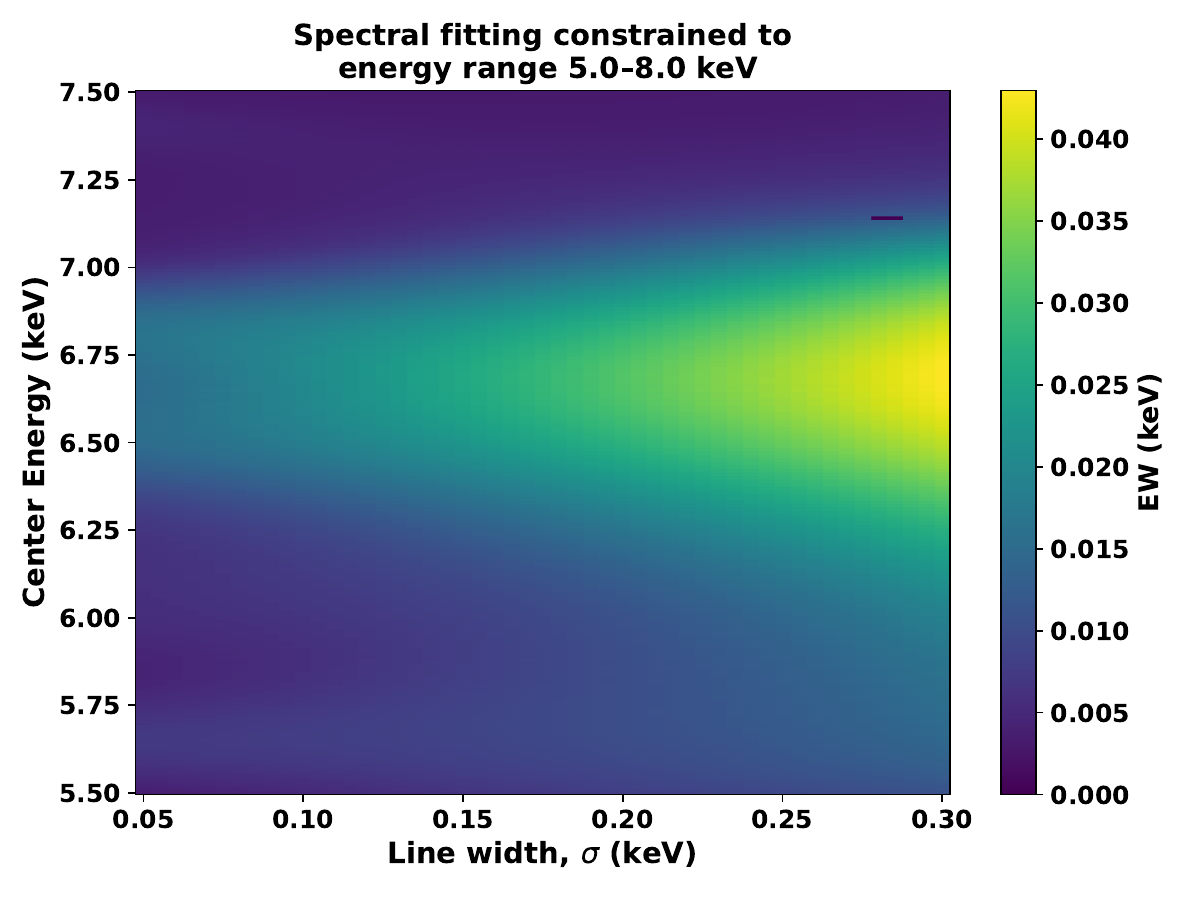}
    \includegraphics[height=6.8cm, width=0.46\linewidth]{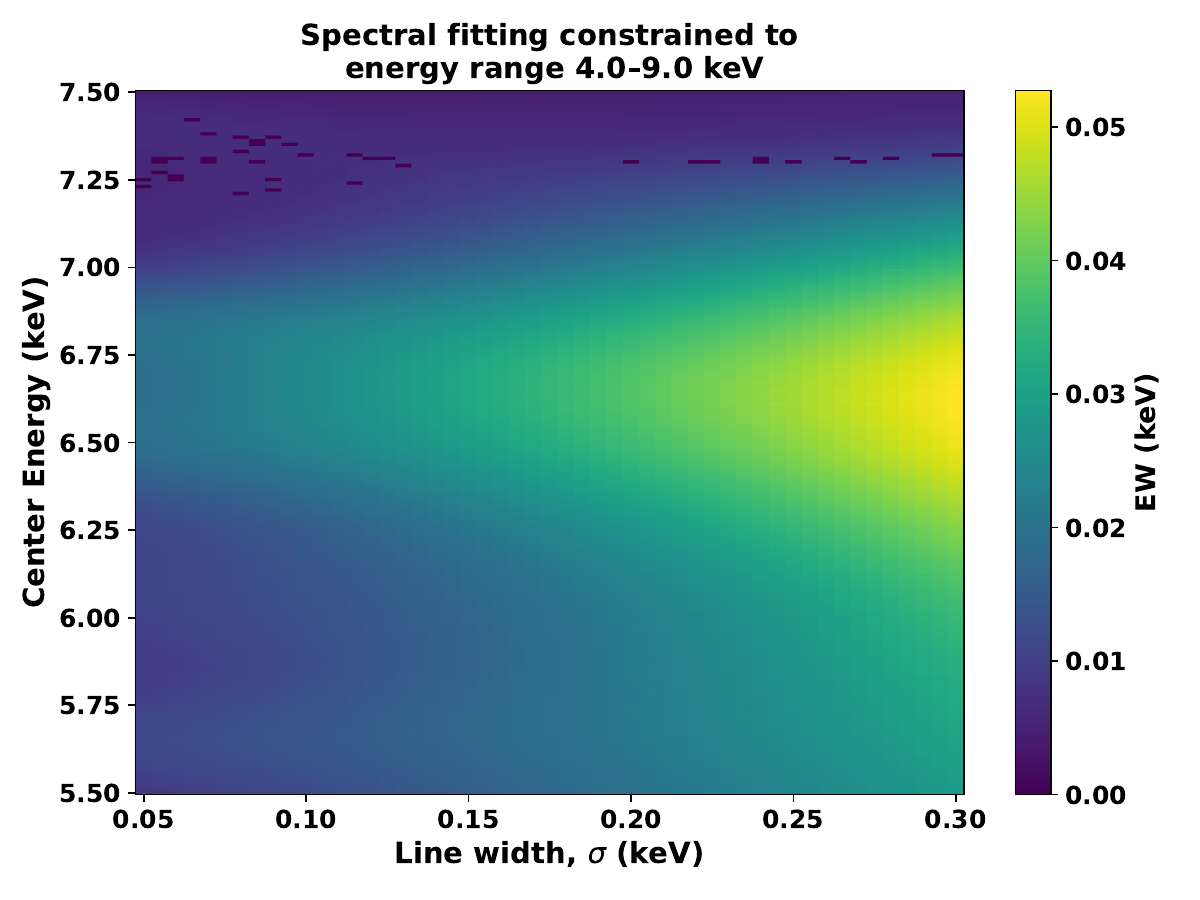}
    \caption{Map of the EW variation across line centroid energy and width, obtained from spectral fits performed in two different fitting energy ranges.}
    \label{fig:EQW_vary_map}
\end{figure*}

We further investigated the detectability of weak Fe K$\alpha$ lines through simulations to validate the upper limits obtained from the combined spectrum. For each of the 12 sources, we used the best-fit model obtained from the combined spectrum of all available observations of that source and added an Fe K$\alpha$ line with a specified EW. Using these models, we simulated the spectra for each source separately. The 12 simulated source spectra were then combined following the same procedure as used for the observed spectra, and the resulting combined spectrum was fitted in the 5.0--8.0 keV energy range to recover the EW of the input Fe K$\alpha$ line. We initially considered an input EW of 30 eV and repeated the simulation for 10,000 independent sets of spectra. The distribution of the recovered EWs is shown in Fig.~\ref{fig:simulated_EQW_validation}. The same procedure was repeated for input EWs of 15 and 20 eV. In all cases, the recovered EWs are very well Gaussian distributed around the respective input values, demonstrating that an Fe K$\alpha$ line with an EW in this range can be recovered from the combined spectra using our analysis procedure. This shows that a line with an EW of this magnitude, if present in the individual source spectra, would be recovered after combining the spectra, and therefore supports the upper limits derived from the combined-spectrum analysis.

To assess whether fixing the line centroid and width could affect the sensitivity of the Fe K$\alpha$ line search, we performed additional fits to the combined spectrum by allowing both parameters to vary. We explored centroid energies in the range 5.5--7.5 keV and line widths between 0.05--0.3 keV, and repeated the analysis over two fitting ranges (4.0--9.0 keV and 5.0--8.0 keV). We find that the upper limits on the EW are highest for centroid energies around $\sim$6.4--6.9 keV, and increase with increasing line width, as expected for broader features (see Fig. \ref{fig:EQW_vary_map}). However, even under these conservative assumptions, the upper limits remain relatively low, with a maximum EW of only $\sim$50 eV. These results are consistent across both fitting ranges and demonstrate that our main conclusion, i.e., the absence of significant Fe K$\alpha$ emission, is robust against reasonable variations in the assumed line parameters.

\begin{table}
 \caption{Best-fit spectral parameters obtained from fitting in the energy range of 5.0--8.0 keV with the \texttt{powerlaw+gauss} model for individual sources and the combined spectrum. The Gaussian line center and line width ($\sigma$) were fixed at 6.4 keV and 0.1 keV, respectively.} 
\label{tab:fit_en_5_8}
\resizebox{\columnwidth}{!}{
\renewcommand{\arraystretch}{1.3}
\hskip-2.0cm\begin{tabular}{cccccc}
\hline
        Source & $\Gamma$ &  $\rm norm $  & $\rm Gauss\,\,norm$  & EW  & $\chi^{2}$/dof \\ 
        & & ($10^{-3}$) & ($10^{-7}$) & (eV) & \\
        \hline
        M51 ULX-7 & $ 1.7 \pm 0.2 $ & $ 0.11_{-0.03}^{+0.05} $ & $< 1.7$ & $< 35.6$ & 248.3/241 \\
        NGC 300 ULX-1 & $ 2.1 \pm 0.1$ & $ 1.8_{-0.4}^{+0.5}$ & $< 14.8$ & $< 40.5$ & 330.9/345 \\
        NGC 1313 X-2 & $ 2.8 \pm 0.1 $ & $ 2.0_{-0.4}^{+0.5} $ & $< 5.2 $ & $< 50.3 $ & 442.8/474 \\
        NGC 5907 ULX-1 & $ 1.8 \pm 0.1 $ & $ 0.22_{-0.05}^{+0.06} $ & $< 3.2$ & $< 44.7$ & 534.5/495 \\
        NGC 7793 P13 & $ 1.8 \pm 0.1 $ & $ 0.6 \pm 0.1 $ & $< 6.6$ & $< 30.3$ & 574.0/555 \\
        Circ ULX-5 & $ 2.5 \pm 0.1 $ & $ 4.2_{-0.7}^{+0.9} $ & $< 11.0$ & $< 29.2$ & 487.4/480 \\
        Ho II X-1 & $ 3.1 \pm 0.2 $ & $ 2.9_{-0.9}^{+1.3} $ & $< 4.1$ & $< 43.4$ & 225.1/254 \\
        Ho IX X-1 & $ 2.4  \pm 0.1 $ & $ 6.0_{-0.9}^{+1.1} $ & $< 15.7$ & $< 22.5$ & 483.1/551 \\
        NGC 1313 X-1 & $ 2.23 \pm 0.07 $ & $ 1.5 \pm 0.2 $ & $< 4.4$ & $< 18.8$ & 630.9/596 \\
        M33 X-8 & $ 2.88 \pm 0.06 $ & $ 12.6_{-1.2}^{+1.3} $ & $< 3.2$ & $< 5.4$ & 544.2/522 \\
        NGC 4559 X-7 & $ 2.8 \pm 0.2 $ & $ 2.4_{-0.7}^{+1.0} $ & $< 8.3$ & $< 60.7$ & 286.6/258 \\
        IC 342 X-2 & $ 1.5 \pm 0.3 $ & $ 0.5_{-0.2}^{+0.3} $ & $< 20.4$ & $< 73.7$ & 99.5/135 \\
        \hline
        Combined & $ 2.36_{-0.03}^{+0.03} $ & $ 1.55_{-0.09}^{+0.10} $ & $< 2.9 $ & $<14.7 $ & 580.7/596 \\
        \hline
        
\end{tabular}
}
\end{table}

\begin{table}
 \caption{Comparison of the upper limits on the EW of the iron line derived from spectral fits across different energy ranges: Range 1 (5.5--7.5 keV), Range 2 (5.0--8.0 keV), and Range 3 (4.0--9.0 keV).} 
 \label{tab:fit_all_en}
 \resizebox{\columnwidth}{!}{
 \renewcommand{\arraystretch}{1.2}
 \begin{tabular}{cccc}
        \hline
        Source & Range 1 (eV) & Range 2 (eV) & Range 3 (eV) \\
        \hline
        M51 ULX-7 & $< 20.9$ & $< 35.6$ & $< 28.2$ \\
        NGC 300 ULX-1 & $< 30.8$ & $< 40.5$ & $< 59.6$ \\
        NGC 1313 X-2 & $< 47.8$ & $< 50.3$ & $< 45.3$ \\
        NGC 5907 ULX-1 & $< 47.9$ & $< 44.7$ & $< 53.2$ \\
        NGC 7793 P13 & $< 20.7$ & $< 30.3$ & $< 42.8$ \\
        Circ ULX-5 & $< 23.4$ & $< 29.2$ & $< 36.2$ \\
        Ho II X-1 & $< 43.9$ & $< 43.4$ & $< 41.5$ \\
        Ho IX X-1 & $< 24.9$ & $< 22.5$ & $< 26.6$ \\
        NGC 1313 X-1 & $< 15.6$ & $< 18.8$ & $< 26.6$ \\
        M33 X-8 & $< 7.8$ & $< 5.4$ & $< 3.6$ \\
        NGC 4559 X-7 & $< 60.2$ & $< 60.7$ & $< 62.0$ \\
        IC 342 X-2 & $< 76.5$ & $< 73.7$ & $< 69.6$ \\
        \hline
        Combined & $< 11.4$ & $< 14.7$ & $< 20.4$ \\
        \hline
         
\end{tabular}
}
\end{table}

\subsection{Hardness Ratio around the eclipse transitions}

\begin{figure*}
    \centering

    \begin{minipage}{0.32\linewidth}
        \centering
        \includegraphics[width=\linewidth]{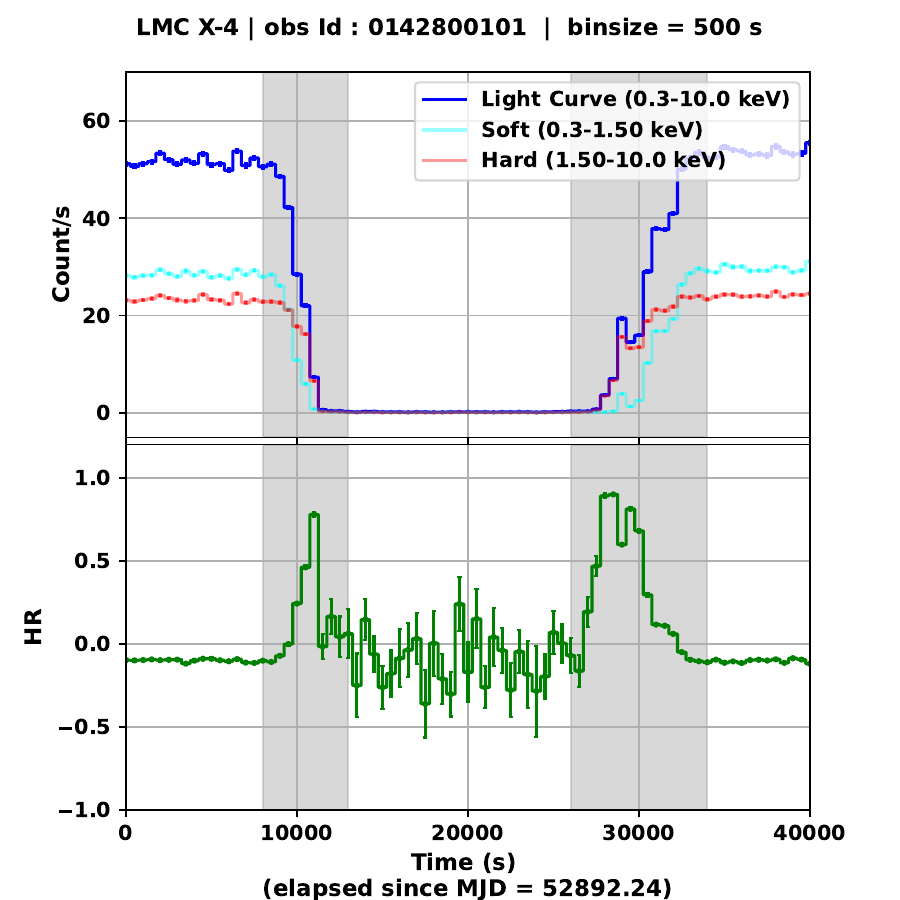}
    \end{minipage}
    \begin{minipage}{0.32\linewidth}
        \centering
        \includegraphics[width=\linewidth]{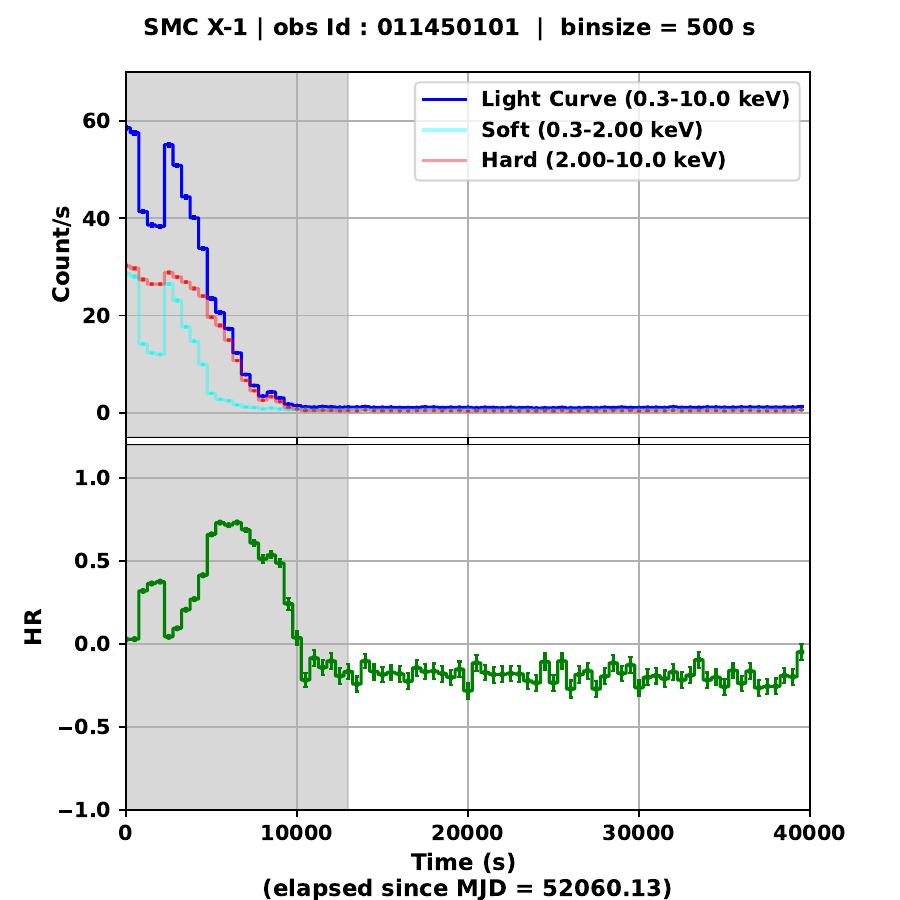}
    \end{minipage}

    \begin{minipage}{0.32\linewidth}
        \centering
        \includegraphics[width=\linewidth]{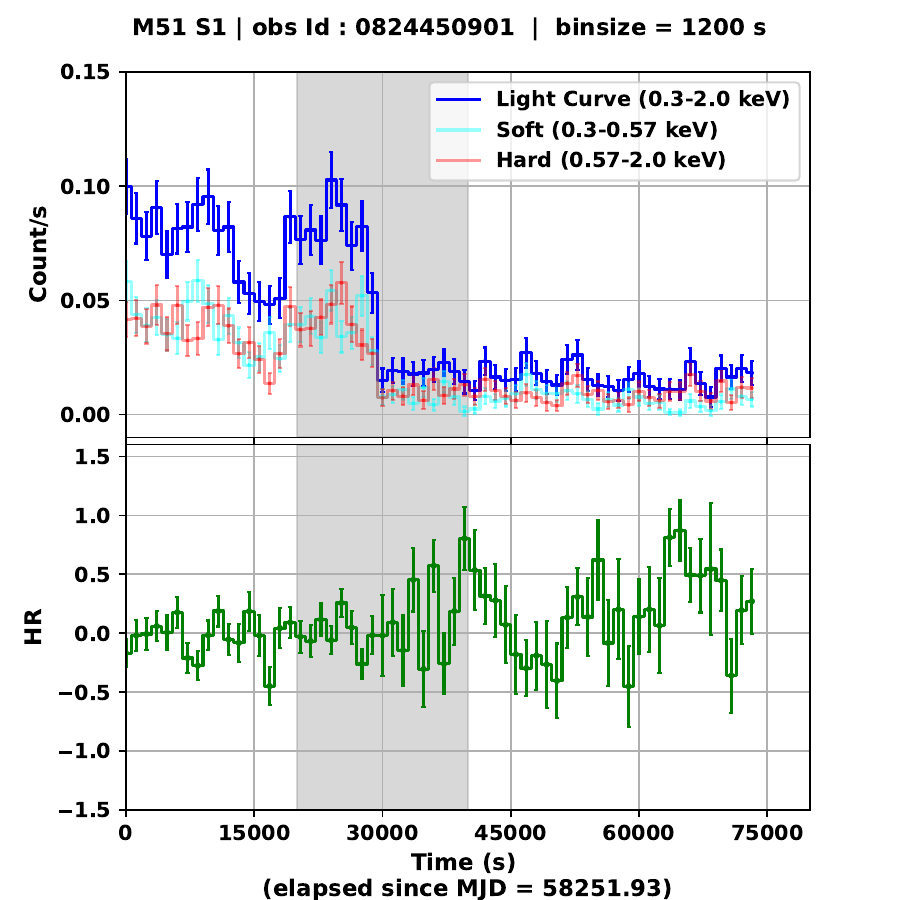}
    \end{minipage}
    \begin{minipage}{0.32\linewidth}
        \centering
        \includegraphics[width=\linewidth]{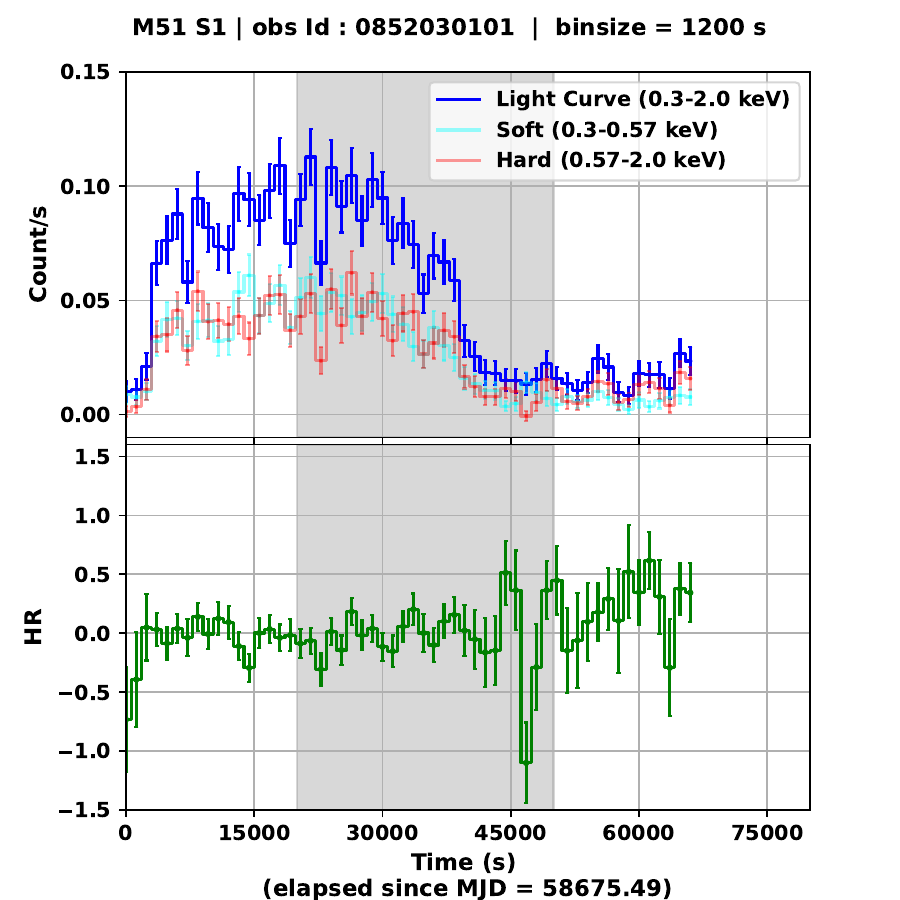}
    \end{minipage}  
    \begin{minipage}{0.32\linewidth}
        \centering
        \includegraphics[width=\linewidth]{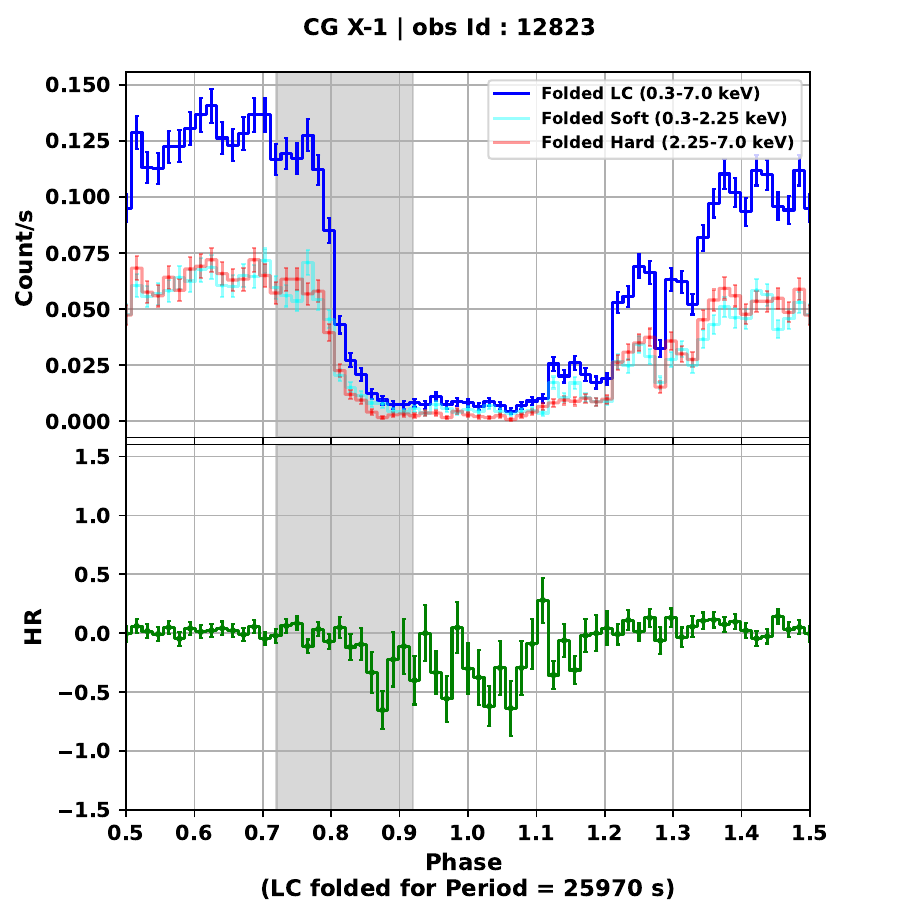}
    \end{minipage}
    
    \begin{minipage}{0.32\linewidth}
        \centering
        \includegraphics[width=\linewidth]{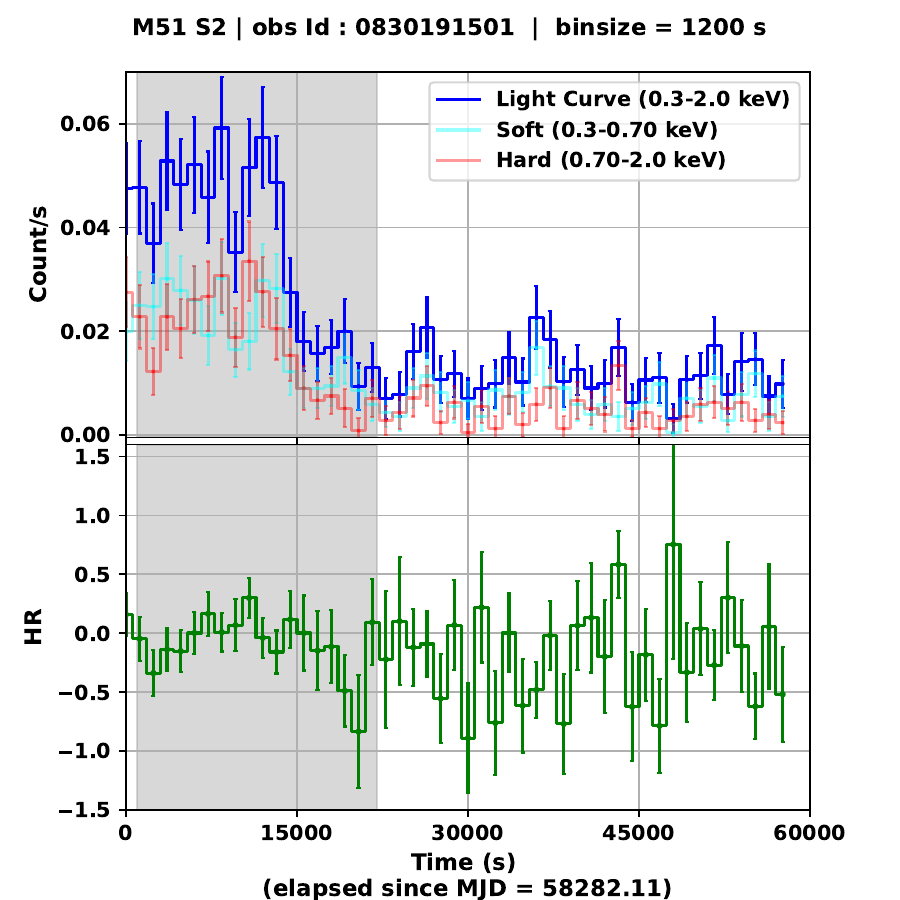}
    \end{minipage}
    \begin{minipage}{0.32\linewidth}
        \centering
        \includegraphics[width=\linewidth]{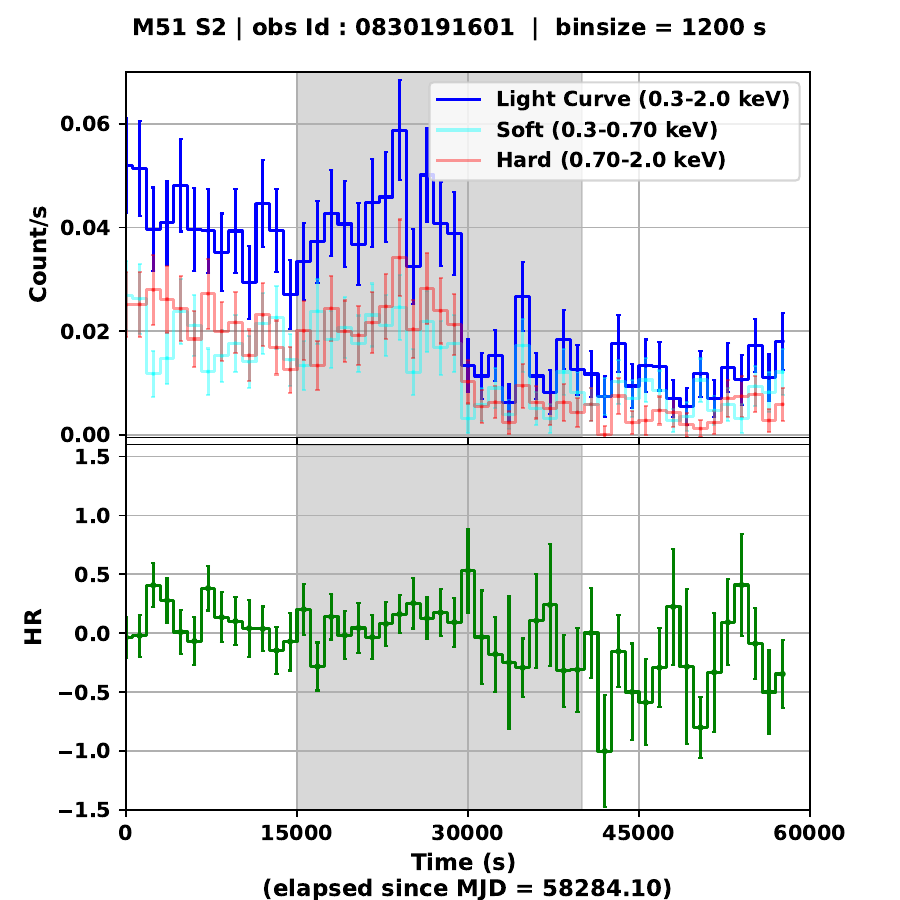}
    \end{minipage}
    \begin{minipage}{0.32\linewidth}
        \centering
        \includegraphics[width=\linewidth]{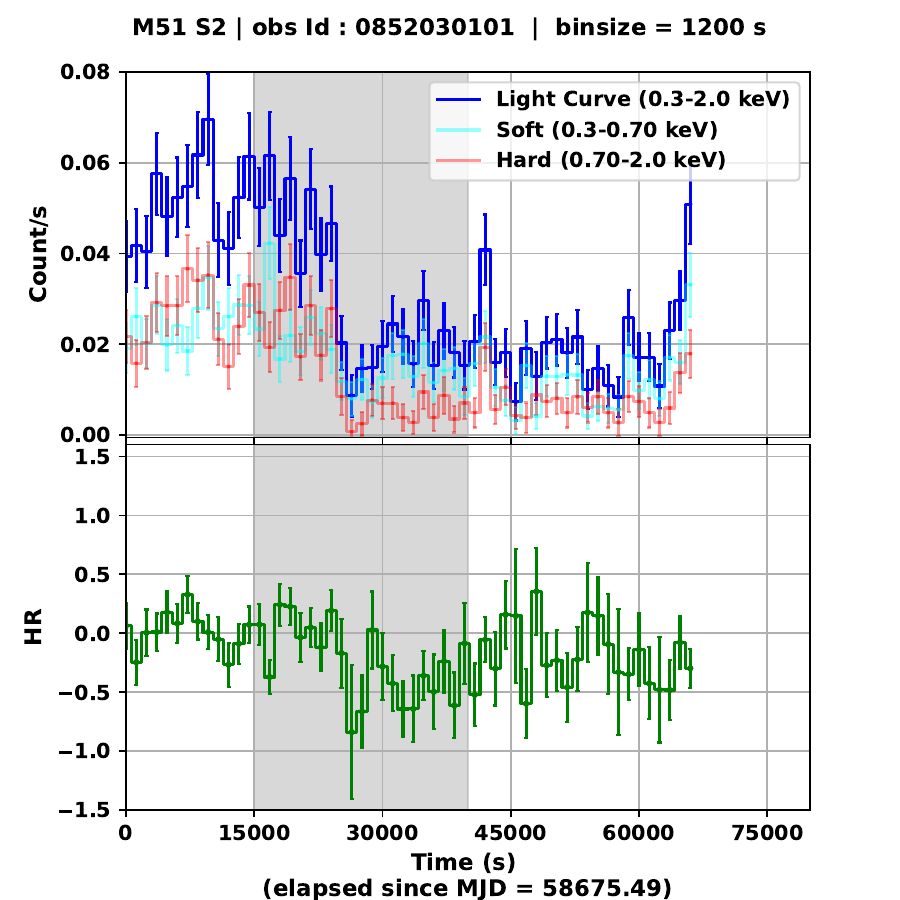}
    \end{minipage}

    \begin{minipage}{0.32\linewidth}
        \centering
        \includegraphics[width=\linewidth]{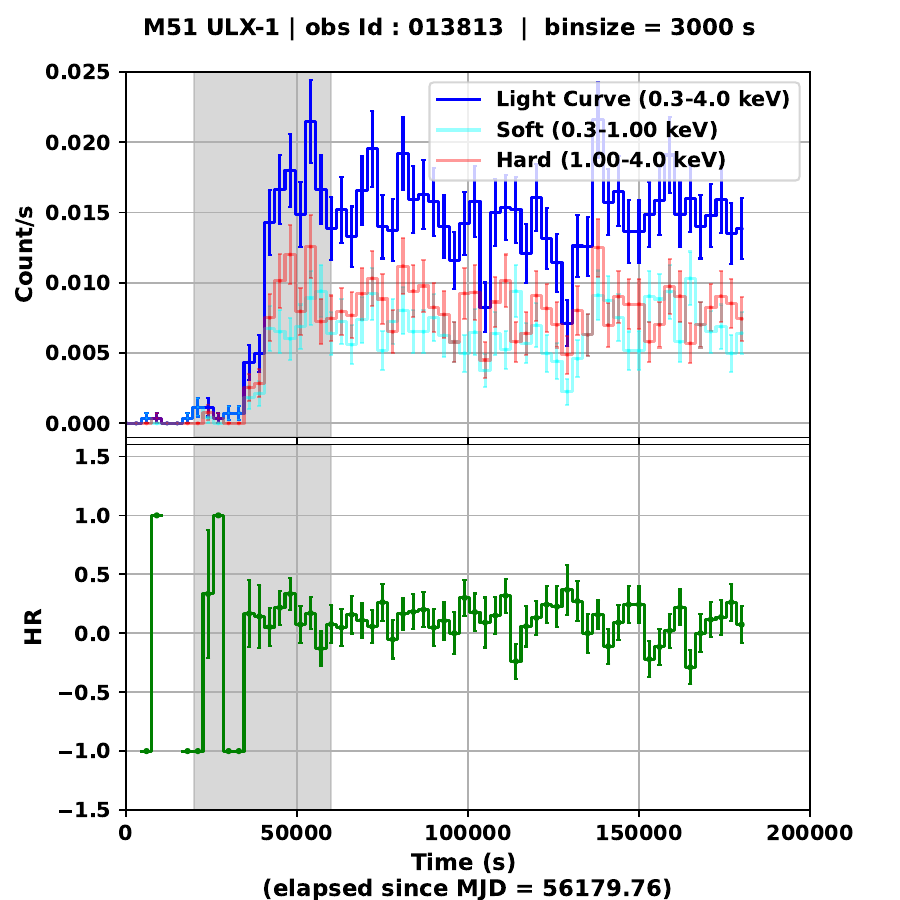}
    \end{minipage}
    \begin{minipage}{0.32\linewidth}
        \centering
        \includegraphics[width=\linewidth]{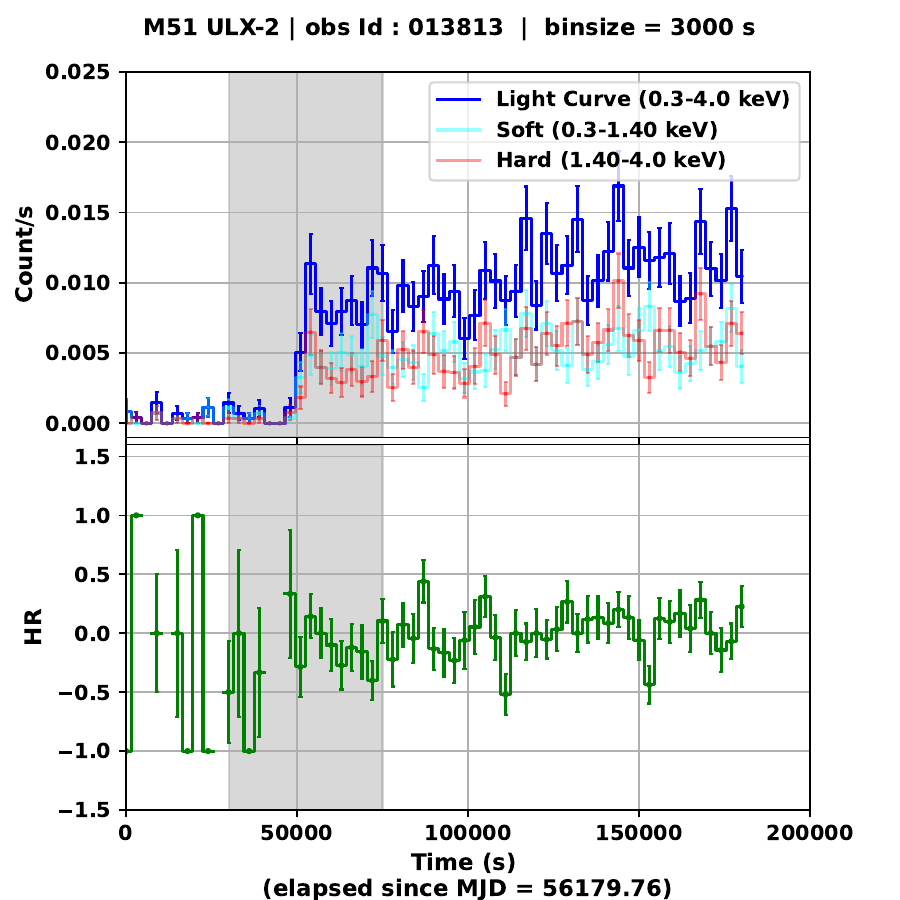}
    \end{minipage}

    \caption{Light curves and hardness ratios around the eclipse transitions. The top row shows results for the well-known HMXBs (LMC X--4 and SMC X--1), while the subsequent rows present the eclipsing ULXs. The grey shaded regions indicate the eclipse transitions.}
    \label{fig:HR with Orbital Phase}
\end{figure*}

We investigated the hardness ratio during the eclipse and the eclipse transitions for five eclipsing ULXs.
The hardness ratio is defined as \( \mathrm{HR} = \frac{H - S}{H + S} \), where \( H \) and \( S \) are the count rates in the hard and soft X-ray energy bands, respectively. To calculate this, we selected the soft and hard X-ray energy bands such that the total counts in both bands are approximately similar. The selected energy bands were: 0.3--0.57 keV and 0.57--2.0 keV for M51 S1, 0.3--0.7 keV and 0.7--2.0 keV for M51 S2, 0.3--2.25 keV and 2.25--7.0 keV for CG X-1, 0.3--1.0 keV and 1.0--4.0 keV for M51 ULX-1, and 0.3--1.4 keV and 1.4--4.0 keV for M51 ULX-2. The eclipse profiles and respective hardness ratios are shown in Fig. \ref{fig:HR with Orbital Phase}. For M51 S1, M51 S2, M51 ULX-1, and M51 ULX-2, only a single eclipse/partial eclipse cycle is covered in the observation, so the hardness ratio is plotted as a function of time. 
In contrast, for CG X-1, the observation spans approximately six orbital cycles \citep{Cg_X1_Qiu_2019}. To examine the hardness ratio around the eclipse transitions, the light curve was folded at the reported orbital period $\rm P=25970\,s$ \citep{Cg_X1_Qiu_2019}. The resulting folded light curve and hardness ratio with orbital phase are also shown in Fig. \ref{fig:HR with Orbital Phase}. 

To compare the hardness ratio behavior with known Galactic eclipsing HMXBs, we selected LMC X--4 and SMC X--1 as representative systems, as they are among the most luminous HMXBs exhibiting sharp eclipses. We used \xmm\ observations with ObsIDs 0142800101 and 011450101, respectively, and reprocessed the data following standard procedures to generate the light curves and hardness ratios (see Fig.~\ref{fig:HR with Orbital Phase}). For LMC X--4, the energy ranges of 0.3--1.5 keV and 1.5--10.0 keV were selected as the soft and hard bands, respectively, while for SMC X--1, the corresponding ranges were 0.3--2.0 keV and 2.0--10.0 keV.

As shown in Fig. \ref{fig:HR with Orbital Phase}, both LMC X--4 and SMC X--1 exhibit a pronounced increase in hardness ratio during eclipse ingress and egress, consistent with the expected additional absorption as the compact object moves behind the stellar companion. In contrast, no significant changes in hardness ratio are observed for the ULXs during eclipse transitions, suggesting that their spectral evolution across eclipses differs fundamentally from that of typical HMXBs. 

\subsection{Flux Ratio during Eclipse to Out-of-Eclipse Phase}

During X-ray eclipses, most X-ray binaries exhibit residual X-ray emission, which is generally attributed to reprocessed emission from the stellar wind \citep{Nafisa_HMXB_eclipsing, Aftab_LMXB, Ketan_2024_eclipse_flare}. The ratio of OOE to eclipse flux shows a wide range in Galactic binaries, typically between $\sim 8-237$ for HMXBs, and $\sim 6-54$ for LMXBs. It is therefore of particular interest to examine this flux ratio for ULXs, where mass accretion rates are expected to be much higher and the accretion geometry potentially more complex.

\begin{figure*}
    \centering
    \begin{minipage}{0.32\linewidth}
        \centering
        \includegraphics[width=\linewidth]{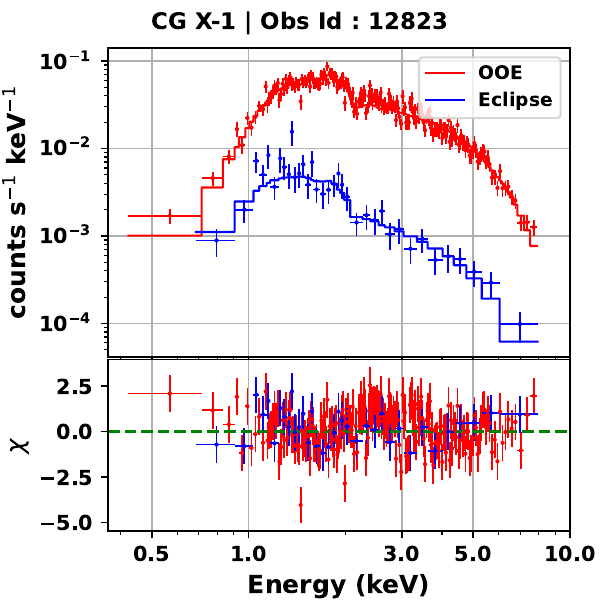}
    \end{minipage}
    \begin{minipage}{0.32\linewidth}
        \centering
        \includegraphics[width=\linewidth]{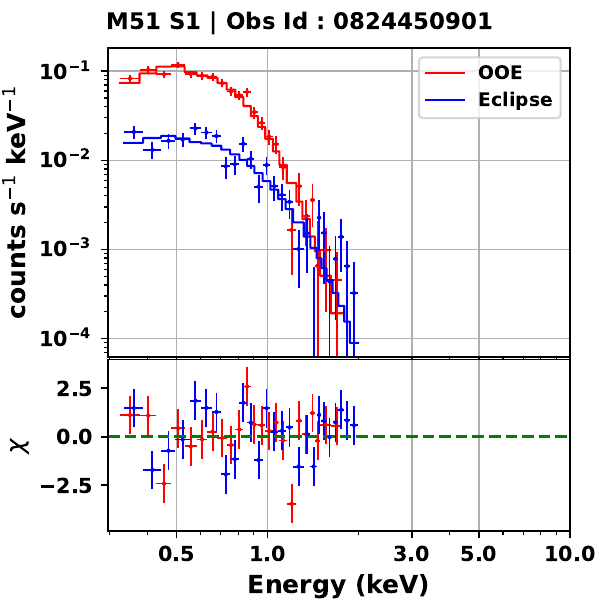}
    \end{minipage}
    \begin{minipage}{0.32\linewidth}
        \centering
        \includegraphics[width=\linewidth]{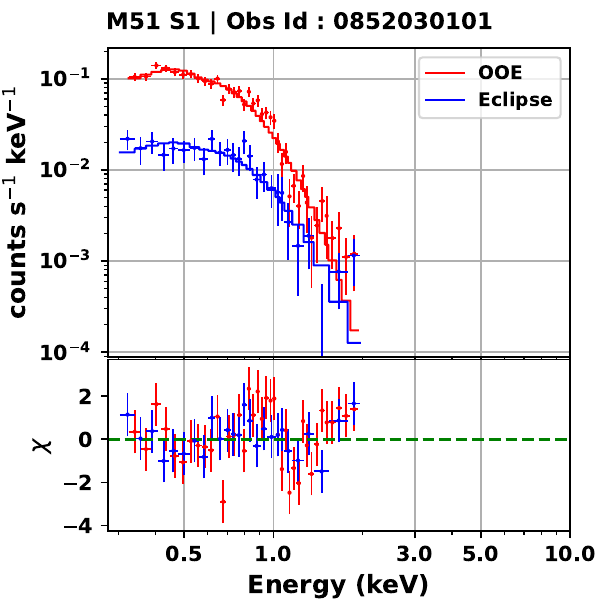}
    \end{minipage}

    \vspace{0.2em} 

    \begin{minipage}{0.32\linewidth}
        \centering
        \includegraphics[width=\linewidth]{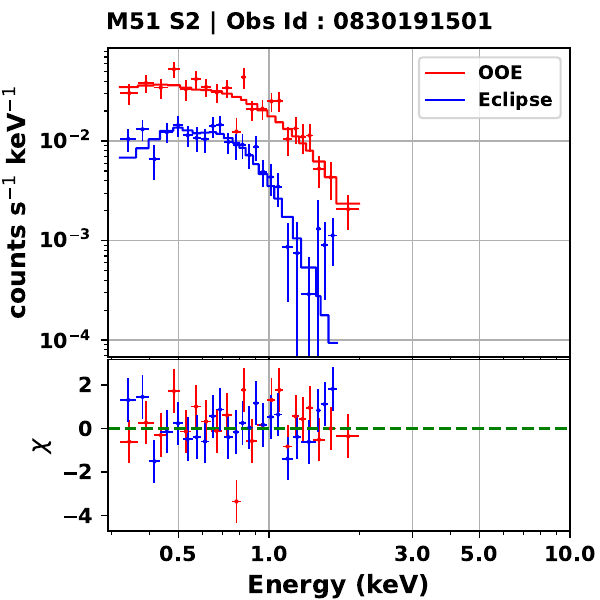}
    \end{minipage}
    \begin{minipage}{0.32\linewidth}
        \centering
        \includegraphics[width=\linewidth]{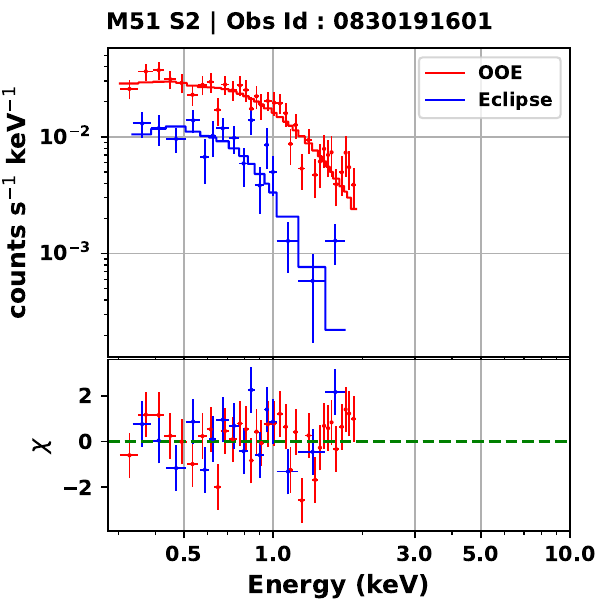}
    \end{minipage}
    \begin{minipage}{0.32\linewidth}
        \centering
        \includegraphics[width=\linewidth]{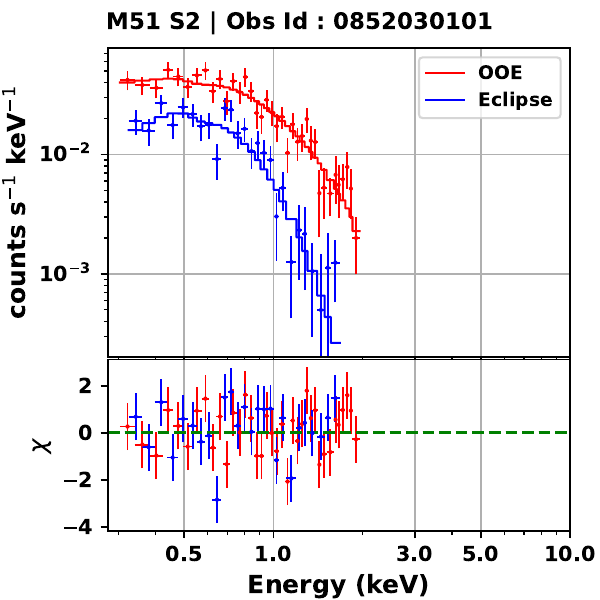}
    \end{minipage}

    \begin{minipage}{0.32\linewidth}
        \centering
        \includegraphics[width=\linewidth]{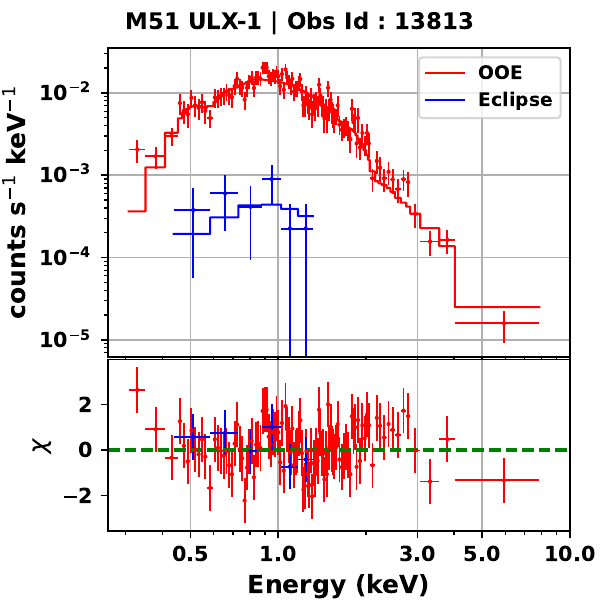}
    \end{minipage}
    \begin{minipage}{0.32\linewidth}
        \centering
        \includegraphics[width=\linewidth]{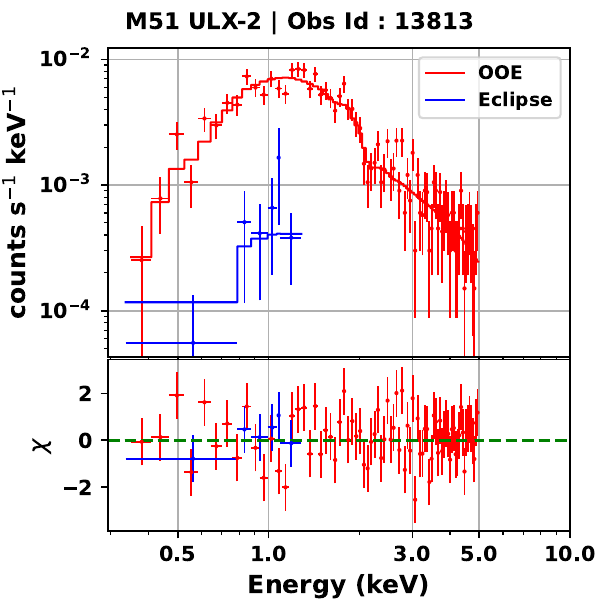}
    \end{minipage}

    \caption{Spectrum during eclipse and OOE phase for the CG X1 (Chandra), M51 S1 (\xmm), M51 S2 (\xmm), M51 ULX-1 (Chandra) and M51 ULX-2 (Chandra). Respective bottom panels show the best-fit residuals.}
    \label{fig:eclipse_spectrum}
\end{figure*}

To calculate the flux, the spectra are fitted using the \texttt{xspec} model \texttt{tbabs*(diskbb)} for M51 S1 and S2, and \texttt{tbabs*(powerlaw)} for CG X-1, M51 ULX-1, and M51 ULX-2. The results are summarized in Table \ref{tab:fit_eclipse_flux_ratio}, and the corresponding spectrum with residuals are shown in Fig. \ref{fig:eclipse_spectrum}. The flux is calculated in the energy range 0.3--2.0 keV for M51 S1/S2 and 0.3--8.0 keV for CG X-1. 
For M51 ULX-1 and ULX-2, insufficient statistics during eclipse prevented a reliable spectral fit. In these cases, we estimated the eclipse flux by first fitting the OOE spectrum and then refitting the eclipse spectrum with all spectral parameters fixed except for the normalization of the powerlaw component. Since no significant counts were detected above 1.4 keV during eclipse, fluxes were computed in the 0.3--1.4 keV energy range.

As indicated in Table \ref{tab:fit_eclipse_flux_ratio}, the flux drops approximately 3--5 times while transitioning from OOE to eclipse phase for M51 S1/S2 and $\sim$23 times for CG X-1. For M51 ULX-1 and M51 ULX-2, the OOE-to-eclipse flux ratios are estimated to be in the ranges $\sim$13.5--52.3 and $\sim$8.1--27.3, respectively, although these values are limited by the available statistics. When compared to most Galactic HMXBs, where flux ratios typically span $\sim$8--237 \citep{Nafisa_HMXB_eclipsing}, the ULXs generally exhibit equal or lower OOE-to-eclipse flux ratios. This suggests that the reprocessing efficiency (eclipse-to-OOE flux ratio) is larger for the ULXs as compared to the Galactic HMXBs.

\begin{table*}
  \caption{Best-fit spectral parameters for selected eclipsing ULXs, comparing eclipse and OOE spectra. For CG X1, M51 ULX-1 and M51 ULX-2, the spectra were fitted with the \texttt{tbabs*(powerlaw)} model, 
whereas for M51 S1 and M51 S2, the \texttt{tbabs*(diskbb)} model was used.}
  \label{tab:fit_eclipse_flux_ratio}
  \fontsize{10}{10}\selectfont
  \resizebox{\linewidth}{!}{
    \renewcommand{\arraystretch}{1.3}
    \hskip-3.0cm\begin{tabular}{ccccccccccc}
      \hline
      \hline
       & &  & \texttt{tbabs} & \multicolumn{2}{|c|}{\texttt{powerlaw}} & \multicolumn{2}{c|}{\texttt{diskbb}} &  & Flux $^{*}$ & Flux Ratio \\
      Source & Obs Id& Phase & $N_{\mathrm{H}}$ ($10^{22}$ cm$^{-2}$) & $\Gamma$ & norm$_\texttt{po}$ ($10^{-5}$) & $kT$ (keV) & norm$_\texttt{diskbb}$ & $\chi^2$/dof & ($10^{-14}$ erg cm$^{-2}$ s$^{-1}$) & (OOE/Eclipse) \\
      \hline
      \hline
      CG X-1 & 12823 & OOE & $ 1.5\pm0.1 $ & $1.68_{-0.07}^{+0.08} $ & $51 \pm 5 $ & - & - & 228.6/197 & $183 \pm 5 $ & $23_{-4}^{+3}$ \\
       &  & Eclipse & $1.0_{-0.2}^{+0.3}$ & $2.1_{-0.3}^{+0.4}$ & $3_{-1}^{+2}$ & - & - & 30.4/33 & $8 \pm 1$ & \\
      \hline
      M51 S1 & 0824450901 & OOE & $0.19_{-0.05}^{+0.06}$ & - & - & $0.13 \pm 0.01 $ & $191_{-103}^{+270}$ & 32.2/20 & $9.0 \pm 0.5 $ & $ 4.5 \pm 0.5 $ \\
       &  & Eclipse & $<0.17$ & - & - & $0.19 \pm 0.04$ & $<8.5$ & 35.9/23 & $2.0 \pm 0.2 $ &  \\
       & & & & & & & & & & \\
       & 0852030101 & OOE & $0.09 \pm 0.03$ & - & - & $0.16 \pm 0.01 $ & $36_{-15}^{+28}$ & 65.5/36 & $10.4 \pm 0.5 $ & $ 5.2 \pm 0.8 $ \\
       &  & Eclipse & $<0.20$ & - & - & $0.19 \pm 0.05$ & $<15.4$ & 16.1/23 & $2.0 \pm 0.3 $ & \\
      \hline
      M51 S2 & 0830191501 & OOE & $<0.06$ & - & - & $0.32 \pm 0.05 $ & $0.4_{-0.2}^{+0.5}$ & 27.2/19 & $6.1 \pm 0.6 $ & $ 4.4 \pm 0.8 $ \\
       &  & Eclipse & $0.2_{-0.1}^{+0.2}$ & - & - & $0.14 \pm 0.04 $ & $<409$ & 18.1/21 & $1.4 \pm 0.2 $ & \\
       & & & & & & & & & & \\
       & 0830191601 & OOE & $<0.03$ & - & - & $0.37 \pm 0.04 $ & $0.18_{-0.06}^{+0.14}$ & 31.6/31 & $5.4 \pm 0.4 $ & $ 3.9_{-0.9}^{+0.6}$ \\
       &  & Eclipse & $<0.28$ & - & - & $0.18 \pm 0.06 $ & $<36.5$ & 20.8/13 & $1.4_{-0.2}^{+0.3}$ & \\
       & & & & & & & & & \\
       & 0852030101 & OOE & $<0.04$ & - & - & $0.33 \pm 0.04 $ & $0.4_{-0.1}^{+0.3}$ & 32.8/34 & $6.4_{-0.5}^{+0.4}$ & $ 3.0_{-0.4}^{+0.3} $ \\
       &  & Eclipse & $0.11_{-0.08}^{+0.11}$ & - & - & $0.17_{-0.03}^{+0.04} $ & $<32.6$ & 30.4/23 & $2.1 \pm 0.2 $ &  \\
      \hline
      M51 ULX-1 & 13813 & OOE & $ 0.36\pm0.05 $ & $3.8 \pm 0.2 $ & $6.5_{-0.7}^{+0.9}$ & - & - & 105.4/106 & $5.6 \pm 0.2 $ & $ 33 \pm 19 $ \\
       &  & Eclipse & $0.36 \, ^{\rm a}$ & $3.81 \, ^{\rm a}$ & $ 0.2 \pm 0.1 $ & - & - & 2.65/5 & $ 0.2 \pm 0.1 $ & \\
      \hline
      M51 ULX-2 & 13813 & OOE & $ 0.33_{-0.07}^{+0.08} $ & $2.2 \pm 0.2 $ & $3.1_{-0.4}^{+0.5}$ & - & - & 81.7/76 & $2.3 \pm 0.1 $ & $ 18 \pm 10 $ \\
       &  & Eclipse & $0.33 \, ^{\rm a}$ & $2.22 \, ^{\rm a}$ & $ 0.2 \pm 0.1 $ & - & - & 2.29/5 & $ 0.13 \pm 0.07 $ & \\
      \hline
      \multicolumn{9}{l}{$\rm ^{*}$: Flux is calculated for the energy range 0.3-8.0 keV for CG X-1, 0.3-2.0 keV for M51 S1/S2, and 0.3-1.4 keV for M51 ULX-1/ULX-2.} \\ 
      \multicolumn{9}{l}{$\rm ^{a}$: parameters are frozen to corresponding values during the fit.} \\ 
        
      
    \end{tabular}
  }
\end{table*}

\section{Discussion}

X-ray absorption/reprocessing is a well-known and well-studied phenomenon in X-ray binaries. There are many characteristics of the X-ray emission from which we can know about the reprocessed emission of X-ray binaries: emission/absorption lines in X-ray spectra, the spectral and timing properties during the eclipse or ingress/egress phases for eclipsing X-ray binaries, etc.
One important outcome of X-ray reprocessing is the production of the Fe K$\alpha$ line, which is the fluorescent emission from the material in the environment of the compact object when illuminated by its X-ray emission.
Both HMXBs and LMXBs clearly show the iron line emission having EW varying from a few eV to $\approx1.7$ keV \citep{Torrejon_2010_iron_line, C_Ng_LMXB, Pradhan_SFXT_SGHMXB, garcia_xmm_newton_Fe_k_alpha_HMXB}. Another manifestation of the reprocessing is an increase in the hardness ratio for eclipsing X-ray binaries during the ingress and egress in Galactic eclipsing X-ray binaries \citep{eclipsing_LMXB_X2127+119_Ioannou, Eclipsing_LMXB_XTE_J1710-281_Gayathri, Vela_X1_HMXB_Abalo_2024, Ajith_Cen_X3, Tamang_Vela_X1}, attributed to the absorption of softer photons by the dense stellar wind near the surface of the companion star. Furthermore, during the eclipse phase, most of the observed X-ray emission is secondary emission reprocessed by a region larger than the compact object.

Nominally, one would expect similar reprocessing phenomena in the ULXs because they are also compact objects powered by accretion from a binary companion star \citep{Rodríguez_Castillo_2020, Qiu2021, 5907_ulx1_orbit_belfiore}. However, signatures of X-ray reprocessing in ULXs have been meagre, except for the cases of absorption lines in the winds \citep{Pinto_blueshifted, Walton_2016, kosec_blueshifted}. 
Only a few ULXs have shown tentative detections of iron emission lines. In M82 X-1, a weak and broadened iron line with an EW of $\rm 59^{+25}_{-27} \, eV$ has been reported \citep{Caballero_M82_X1}. A possible detection of an iron emission line has also been reported in NGC 7456 ULX-1, with an EW of $2.0^{+1.5}_{-1.1}$ keV \citep{Monda_NGC7456_ULX1}, which is significantly higher than that observed in M82 X-1. Recently, a weak iron emission feature has been reported from the ULXP M82 X-2 based on XRISM observations \citep{2026_kobayashi_XRISM_M82_X2}. Additionally, an iron emission line with an EW of $\rm 343 \pm 93 \, eV$ was reported from the ULXP NGC 300 ULX-1 only during the 2010 observation \citep{Carpano}; however, no pulsations were detected during that observation. Overall, iron emission lines remain largely absent in most ULXs.

To look more into the reprocessed emission from ULXs, we have carried out the following three types of investigations: (i) search for the iron $\rm K\alpha$ emission line feature, (ii) hardness ratio evolution during ingress and egress, and (iii) flux ratio between OOE and eclipse phases. We have investigated a large volume of data carefully selected from all existing XMM-Newton and Chandra databases. For studying the iron $\rm K\alpha$ emission line, we selected all \xmm\ observations of seven bright, well-known ULXs and five ULXPs. For studying the hardness ratio during ingress/egress of eclipse and flux ratio during OOE and eclipse phase, we have selected known eclipsing ULXs: CG X-1, M51 S1, M51 S2, M51 ULX-1, and M51 ULX-2. Since M51 S1 and M51 S2 are well resolved with \xmm\ instrument, we used \xmm\ observations for these two sources, while Chandra data were utilized for the sources CG X-1, M51 ULX-1, and M51 ULX-2.

In our analysis, we did not detect any significant iron K$\alpha$ line feature. For the individual sources, the 90\% confidence upper limit on the EW lies in the range 15--60 eV, whereas for the combined spectra we obtained a much tighter constraint of 11--20 eV, which is a significant improvement over previous studies by a factor of a few.
For instance, \cite{NGC_1313_X1_Ho_IX_X1_walton} obtained the upper limit on the iron $\rm K\alpha$ line of $\sim$30 eV for Ho IX X-1 and $\sim$50 eV for NGC 1313 X-1. In comparison, Galactic X-ray binaries almost ubiquitously show iron K$\alpha$ emission in the majority of sources, with EWs varying from a few eV to keV depending on the system type. 

\begin{figure}
	\includegraphics[ height=8cm, width=\columnwidth]{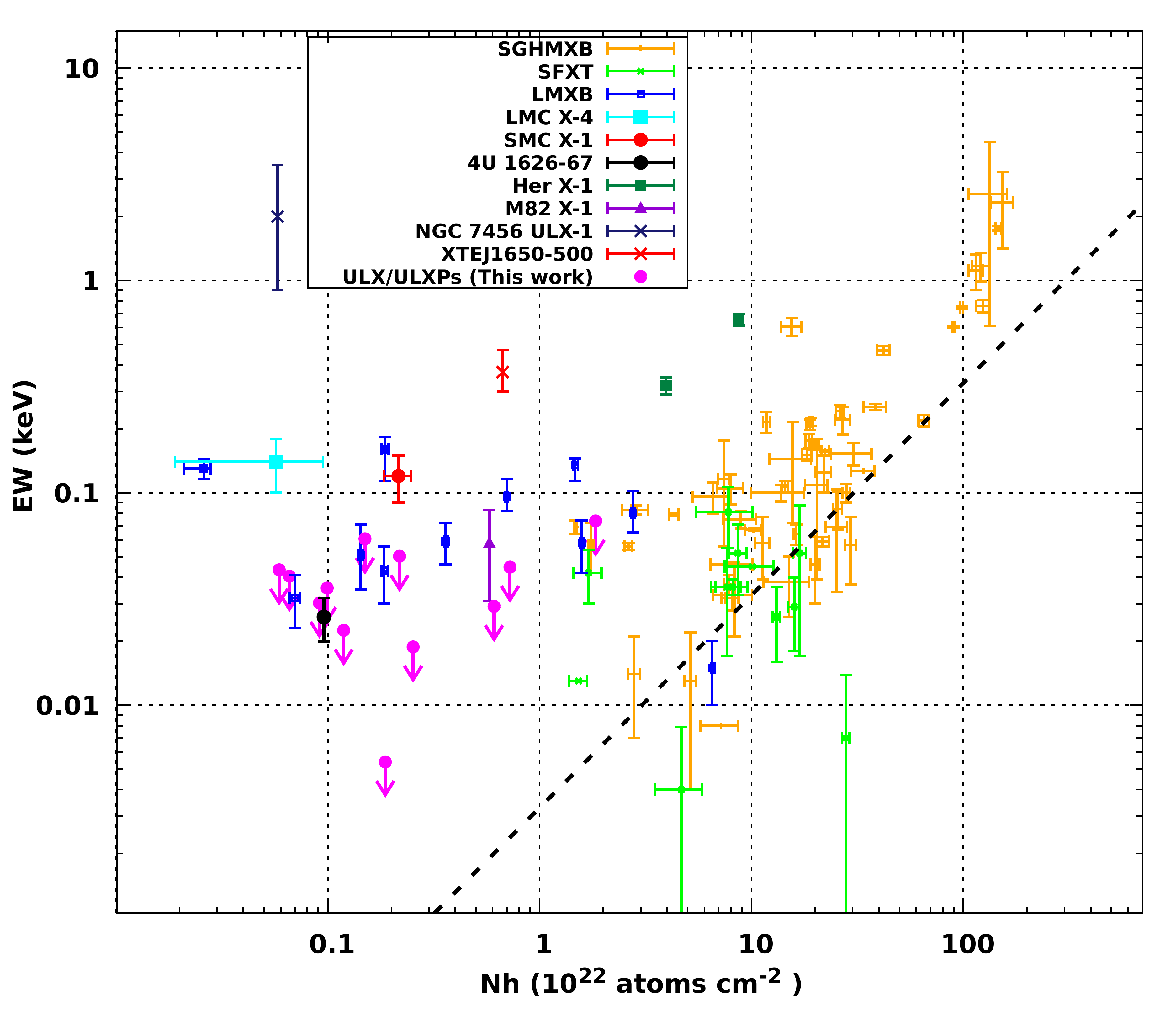}
    \caption{Comparative studies of hydrogen column density ($N_H$: number of hydrogen atoms cm$^{-2}$ along the line of sight) and EW of iron K$\alpha$ line for SgHMXBs, SFXTs  \citep{Pradhan_SFXT_SGHMXB}, LMXBs \citep{C_Ng_LMXB}, LMC X-4, SMC X-1 \citep{Paul2002}, Her X-1 \citep{naik_her_x-1}, 4U 1626-67 \citep{Sharma2023b}, M82 X-1 \citep{Caballero_M82_X1}, NGC 7456 ULX-1 \citep{Monda_NGC7456_ULX1}, XTE J1650-500 \citep{Walton_XTEJ1650_500} and ULXPs. 
    The column density values are taken from previous studies mentioned in Table \ref{tab:total_GTE}. The black dashed line corresponds to the expected EW value for a given column density ($\rm EW = 0.00329 \, N_{H}^{22} \, keV $) assuming the spherical reprocessing geometry around the X--ray source \citep{Torrejon_2010_iron_line}. }
    \label{Nh_vs_EQW}
\end{figure}

Next, we constructed a curve of growth for the ULXs by utilizing the EW values derived from this work (Table \ref{tab:fit_en_5_8}) and column density values from the previous literature (Table \ref{tab:total_GTE}). The resulting relation is shown in Fig.~\ref{Nh_vs_EQW}. The dashed curve represents the model prediction for the relation between the EW of the iron line and the column density ($N_{\mathrm{H}}$) for a spherically symmetric stellar wind distribution \citep{Torrejon_2010_iron_line}. 
For comparison, we also included measurements from classical SG HMXBs, SFXTs \citep{Pradhan_SFXT_SGHMXB}, LMXBs \citep{C_Ng_LMXB}, accreting disk-fed X-ray pulsars \citep{Paul2002, naik_her_x-1, Sharma2023b}, the Galactic stellar-mass black hole binary XTE J1650–500 \citep{Walton_XTEJ1650_500}, and the two ULXs with previously reported iron lines, M82 X-1 \citep{Caballero_M82_X1} and NGC 7456 ULX-1 \citep{Monda_NGC7456_ULX1}. Fig. \ref{Nh_vs_EQW} clearly demonstrates that ULXs occupy the low $N_{\rm H}$ regime compared to Galactic binaries. It is somewhat counterintuitive that despite their extreme luminosities and expected presence of dense reprocessing material, ULXs consistently exhibit relatively low absorbing column densities.


We also estimated the expected absorption in the ULX binary environment assuming a spherically symmetric radiation-driven wind from the companion star, following the model given by \cite{Radiation_driven_wind_model_Castor_1975}. For this, we adopted binary parameters from previous studies, in a manner similar to \cite{Ajith_Cen_X3} and \cite{Tamang_Vela_X1} for Galactic HMXBs. The wind velocity profile is given by:

\begin{equation}
v(r) = v_{\infty} \left( 1 - \frac{R_*}{r} \right)^{\beta}
\end{equation}

where $v_{\infty}$ is the terminal wind velocity, $R_*$ is the companion star radius, $\beta$ velocity gradient parameter fixed to a value 0.8 \citep{Friend_Abbott_1986}. Assuming the donor star is losing mass isotropically with a mass loss rate $\dot{M}$, the radial mass density profile of the wind is given by:

\begin{equation}
\rho(r) = \frac{\dot{M}}{4\pi r^2 v(r)} = \frac{\dot{M}}{4\pi r^2 v_{\infty}} \frac{1}{\left(1 - \frac{R_*}{r}\right)^{\beta}}
\end{equation}

\begin{figure}
	\includegraphics[ height=7.5cm, width=\columnwidth]{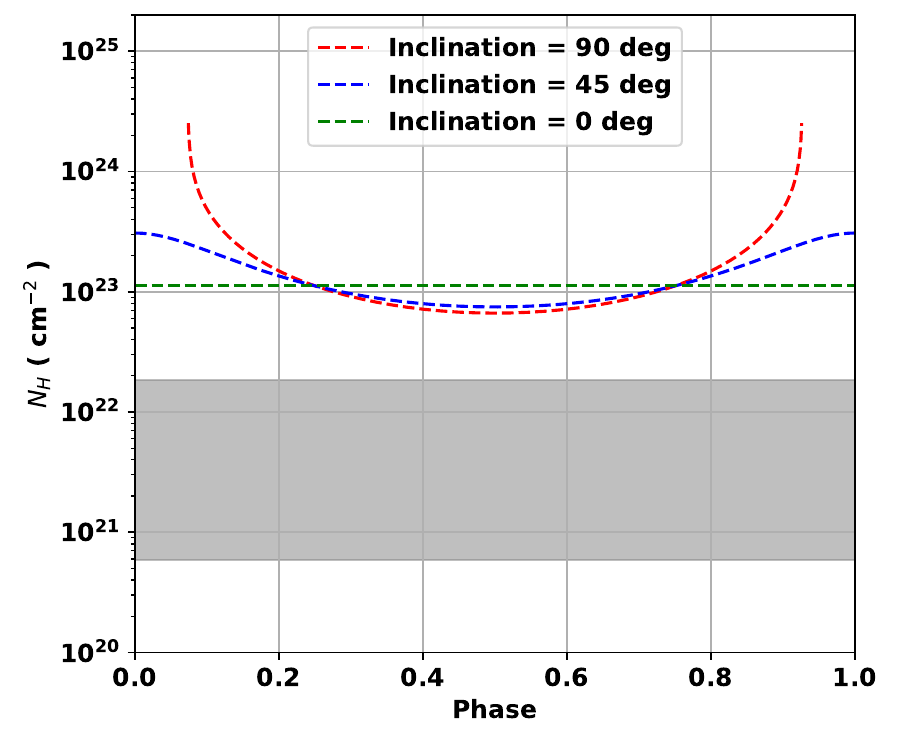}
    \caption{Expected column density calculations assuming accretion through a spherically symmetric wind around the companion. The horizontal gray region denotes the range for measured column densities reported in previous studies (Table \ref{tab:total_GTE}) for ULXPs/ULX sources.}
    \label{fig:N_h_calculation_istotropic_wind}
\end{figure}

The hydrogen mass density is $\rm \rho_H(r) = x_H \, \rho(r) $, $x_H$ is the hydrogen mass fraction (adopted as $x_H \approx 0.7$ from \citep{Wilms2000}). The equivalent hydrogen column density is calculated by integrating the above equation along the line of sight.

\begin{equation}
N_H(\phi') = \kappa \int_{x_{\phi'}}^{\infty} \rho_H(r(x',y_{\phi'},z_{\phi'})) \, dx'
\end{equation}

where $\phi'$ is the orbital phase, $\kappa$ is the conversion factor for mass density to particle density, $r$ is the radial distance from the donor star center, and $x'$ is the line-of-sight coordinate. We adopted the binary parameter values for the ULXP M82 X-2 reported by \citet{M82_X2_orbital_decay_Bachetti}. 
Assuming a circular binary orbit of radius $R_{\rm orbit} = 18.0 R_{\odot}$ with the terminal velocity $ v_{\infty} = 1500 $ km/s, the companion star radius $R_* = 8.0 R_{\odot}$, and the mass loos rate from the companion star is $\dot{M} = 4.7 \times 10^{-6} \, M_{\odot} \rm yr^{-1}$, we calculated the expected column density for viewing the binary at three different angles: edge-on (inclination angle; i = $90^{\rm o}$), face-on (i = $0^{\rm o}$), and an intermediate angle i = $45^{\rm o}$. These results are shown in Fig. \ref{fig:N_h_calculation_istotropic_wind}, where the observed column density ($\rm N_{H}$) values for the ULXs are shaded in gray color. The measured column densities for ULXs are systematically lower than the values predicted from wind accretion onto the compact object. We note that this estimate is intended only as an order-of-magnitude comparison to assess whether reprocessed emission from the stellar wind could contribute to the observed spectra, and to provide a qualitative comparison with the accretion scenario in classical HMXBs. It is not meant to represent a physically accurate description of the accretion environment, which in ULXs is expected to be governed by super-Eddington accretion 
with anisotropic outflows and complex geometry. Nevertheless, even in systems where accretion proceeds primarily via Roche-lobe overflow rather than classical wind-fed accretion, the high-mass companion is expected to drive a stellar wind, and its interaction with the ULX X-ray emission may still contribute to the observed reprocessed component.


Also, we do not observe any significant change in the hardness ratio during eclipse ingress or egress in ULXs (Fig. \ref{fig:HR with Orbital Phase}), whereas eclipsing Galactic X-ray binaries exhibit a clear increase in hardness ratio near these phases. Additionally, the flux drops by approximately 3--5 times while transitioning from OOE to eclipse phase for M51 S1/S2 and $\sim$23 times for CG X-1. For M51 ULX-1 and M51 ULX-2, the OOE-to-eclipse flux ratios are estimated to be $\sim$13.5--52.3 and $\sim$8.1--27.3, respectively, owing to limited statistics. Overall, ULXs appear to exhibit higher eclipse-to-OOE flux compared to Galactic X-ray binaries \citep{Nafisa_HMXB_eclipsing, Aftab_LMXB}. Also, it is intriguing that for M51 S1, the OOE and eclipse spectra almost overlap above 1.3 keV (Fig \ref{fig:eclipse_spectrum}). The eclipsing nature of these sources has been discussed in previous studies \citep{Urquhart2016, M51_S1_S2_Wang, Cg_X1_Qiu_2019}, where the observed eclipse-like features were interpreted as occultation of the central X-ray emitting region by the companion star.

Unlike HMXBs, which show a wide range of $N_{\rm H}$ values and prominent iron emission lines with varying EWs, ULXs typically exhibit lower $N_{\rm H}$ and a lack of detectable iron emission lines. Furthermore, ULXs do not display any significant increase in hardness ratio during eclipse ingress and egress that is commonly observed in HMXBs, due to absorption by dense stellar winds from the companion star. The higher eclipse-to-OOE flux ratio in ULXs suggests the presence of strong scattering. Considering this, the absence of detectable iron emission lines and the lack of increased hardness during eclipse could be indicative of either lower metallicity in the winds of ULX companion stars or a highly ionized stellar wind environment, provided that the observed eclipses are caused by the companion star. However, the lack of significant hardness variation during eclipse ingress and egress may also raise the possibility that structures other than the companion star contribute to the observed obscuration.

Population studies show that metal-poor galaxies with $\rm Z < 0.2~Z_{\odot}$ host a significantly larger number of ULXs compared to metal-rich systems of similar star-formation rate \citep{more_ULX_in_metal_poor_Mapelli_2010}, suggesting that the environments surrounding most ULXs are metal-poor \citep{ULX_optical_pakull_mirioni_2002, ULX_metal_poor_cartwheel_galaxy_Maprlli_2009, Maritza_NGC_925_ULX_optical}. For example, \citet{Maritza_NGC_925_ULX_optical} found that ULX-1 in NGC 925 lies in a region of very low gas metallicity, while \citet{NGC_4559_Optical_Soria_2005} suggested that the ultraluminous source NGC 4559 X-7 also resides in a metal-poor environment. These results reinforce the general trend that ULXs are preferentially associated with low-metallicity surroundings. However, in the same galaxy NGC 925, ULX-3 is located in a region of comparatively higher metallicity \citep{Maritza_NGC_925_ULX_optical}. In addition to metallicity, there is strong evidence for a highly ionized medium around many ULXs, inferred from characteristic emission and absorption lines detected both in X-rays \citep{Kosec_ionized_emission_absorption_2021} and optical \citep{ULX_optical_pakull_mirioni_2002, NGC_1313_X2_Optical_Zampieri_2004, Maritza_NGC_925_ULX_optical}. Taken together, these findings suggest that the surroundings of most ULXs are either metal-poor or dominated by highly ionized gas.

The relatively higher eclipse-to-OOE flux ratio observed in ULXs suggests that a non-negligible fraction of emission persists during eclipse. One possible explanation is a partial-eclipse scenario, in which the X-ray source is only partially occulted. Such a scenario can arise if the emission region is spatially extended, as in accretion disk corona (ADC) sources \citep[e.g., 4U 1822--37,][]{4U_1822_37_Chetana_2010}, or if the line of sight grazes the companion star, leading to partial covering by dense stellar wind material \citep[e.g., IGR J16393--4643,][]{IGR_J16393_4643_sanhita_2020}.
However, the observed eclipse transitions in ULXs are sharp (Fig.~\ref{fig:HR with Orbital Phase}), indicating that the X-ray emitting region is compact. In a partial eclipse scenario, smoother ingress and egress would be expected due to the gradual occultation of an extended emission region or partial covering effects. The sharp transitions observed here, therefore, argue against a partial eclipse interpretation. In fact, the eclipse transitions in ULXs appear sharper than those observed in classical eclipsing HMXBs such as SMC X--1 and LMC X--4, further supporting the conclusion that the X-ray source is compact and fully occulted during eclipse.

While the absence of Fe K$\alpha$ emission in ULXs can be interpreted in terms of lower metallicity in the surrounding medium, supported by the lack of significant hardness ratio variation during ingress/egress and the relatively higher eclipse-to-OOE flux ratio, alternative explanations have also been proposed in the literature based on the nature of the accretion flow in these systems.
ULXs are expected to accrete at super-Eddington rates, leading to the formation of a geometrically thick accretion disc and powerful, radiation-driven outflows. These outflows are highly anisotropic and are thought to form a funnel-like structure along the polar direction, within which the X-ray emission is preferentially beamed \citep{King2023}. In such a geometry, a large fraction of the intrinsic emission escapes without interacting with the surrounding material, thereby reducing the amount of radiation available for reflection and fluorescence.
Consequently, the fraction of emission intercepted and reprocessed by the surrounding medium (e.g., the accretion disc or stellar wind) is significantly reduced compared to standard sub-Eddington systems. This provides a natural explanation for the absence or weakness of Fe K$\alpha$ line emission in ULXs, without invoking low metallicity in the surrounding environment.
In the super-Eddington accretion scenario, the radiation is largely confined within the funnel, which is roughly perpendicular to the orbit. This component, therefore, should not be visible in the case of the eclipsing ULXs, and highly luminous ULXs therefore should not be eclipsing systems. On the other hand, X‐ray emission originating from the outflows and the outer regions of the accretion disc may remain observable for the high inclination systems, provided that the mass accretion rate is sufficiently high. The eclipsing ULXs, with high inclination, should therefore have relatively low observed luminosity and possibly a higher iron line EW. However, in the available Chandra and XMM-Newton data , the eclipsing ULXs are quite faint, and some are detected only up to 2 keV, which do not allow a search for the iron emission line. In addition, if the high-inclination systems have two relatively less luminous visible regions at different distances from the compact object (perhaps the accretion disk and outflow regions), that can explain the observed eclipse behavior of these ULXs. A smaller region near the compact object can cause the sharp eclipse profile, and a relatively larger component (outflow) can still remain visible during the eclipse. However, the currently available data do not allow us to clearly probe these hypotheses.

\section*{Acknowledgements}

We sincerely thank the referee for a very careful reading of the manuscript and for valuable suggestions that have significantly improved it. We also sincerely thank Ajith Balu and Abhisek Tamang for providing code and assistance in calculating the expected column density in the ULX binary environment, assuming a spherically symmetric radiation-driven wind from the companion star. 

This research has made use of data obtained with XMM-Newton, an ESA science mission with instruments and contributions directly funded by ESA Member States. This research also made use of the data obtained with the Chandra X-ray Observatory, a flagship mission of NASA. This research has made use of archival data and software provided by NASA’s High Energy Astrophysics Science Archive Research Center (HEASARC), which is a service of the Astrophysics Science Division at NASA/GSFC.


\section*{Data Availability}

This paper employs a list of Chandra datasets, obtained by the Chandra X-ray Observatory, contained in the Chandra Data Collection~\dataset[DOI: 10.25574/cdc.581]{https://doi.org/10.25574/cdc.581}. Also, the XMM-Newton data used in this work can be accessed through the HEASARC archive at \url{https://heasarc.gsfc.nasa.gov/cgi-bin/W3Browse/w3browse.pl}.

\bibliography{ULX_references}{}
\bibliographystyle{aasjournalv7}


\appendix
\renewcommand{\thetable}{A\arabic{table}} 
\renewcommand{\theHtable}{A\arabic{table}}
\setcounter{table}{0}                     

\renewcommand{\thefigure}{A\arabic{figure}} 
\renewcommand{\theHfigure}{A\arabic{figure}} 
\setcounter{figure}{0}                     
\vspace{-0.8cm}
\section{ULX Sample Selection and Screening Procedure}
\label{apx:selection}

To investigate three key aspects of ULX reprocessed emission: (i) iron line features in the brightest ULXs, (ii) hardness ratio evolution during eclipse transitions, and (iii) flux ratios between eclipse and OOE phases, we constructed dedicated source samples as described below.

For the iron line study, we first considered the catalogue of \cite{Earnshaw_ULX_catalogue}, which provides several flux-related parameters. From these, we selected sources based on the mean flux estimate $S_{\mathrm{Flux,EP5}}$ (4.5--12 keV band), applying a threshold of $S_{\mathrm{Flux,EP5}} > 1.65 \times 10^{-13}\,\mathrm{mW\,m^{-2}}$. We also considered the catalogue of \cite{walton_2022}, selecting sources based on the peak flux $F_{\mathrm{peak,4XMM}}$ (0.2--12 keV band), with a threshold of $F_{\mathrm{peak,4XMM}} > 8.90 \times 10^{-13}\,\mathrm{mW\,m^{-2}}$.
Since these two flux measures represent different quantities, the selection criteria were applied independently to the two catalogues. 
The resulting source lists were then merged, with overlapping entries consolidated into a single final sample for further analysis.
For each selected source, all available \emph{XMM-Newton} observations were retrieved and systematically examined. Observations affected by source confusion, contamination, non-detections, or strong background flaring were excluded following standard screening procedures. 
We further included a few well-established ULX candidates reported in the literature that are not listed in either catalogue, as these meet our selection criteria (outlined below).

From the final source list, cleaned light curves in the 5--8 keV energy band were extracted for all observations. We computed the average count rate, total good time exposure (GTE), and total counts, and selected only sources with total counts $\gtrsim 6000$ for detailed spectral analysis to study the iron K$\alpha$ line. This resulted in a final sample of 12 bright ULXs for the iron line study, listed in Table \ref{tab:final_sample_ULX}. 
The XMM-Newton observations used in the iron K$\alpha$ emission line search for these sources are listed in Table~\ref{tab:observation_log}.

For the analysis of hardness ratios evolution during eclipse transitions and flux ratios between eclipse and out-of-eclipse phases, we selected all five known eclipsing ULXs: CG X-1, CXOM51 J132943.3+471135 (M51 S1), CXOM51 J132946.1+471042 (M51 S2), CXOM51 J132940.0+471237 (M51 ULX-1), and CXOM51 J132939.5+471244 (M51 ULX-2).

\begin{table}[h]
\centering
\caption{Final list of the ULX sample mentioning the total GTE and total counts (in the energy range 5-8 keV).}
\label{tab:final_sample_ULX}
\begin{tabular}{c l l l l c c}
\hline
Sr. No. & SrcID/RecNo.$^{\rm a}$ & IAU Name & Galaxy Name & General Name & GTE (ks) & Counts \\
\hline
1 & 6996 & 3XMM J081928.9+704219 & Holmberg II & Ho II X-1 & 232.64 & 6057 \\
2 & 10388 & 3XMM J034615.7+681112 & IC 0342 & IC 0342 X-2 & 85.39 & 6154 \\
3 & 352633 & 3XMM J123551.7+275604 & NGC 4559 & NGC 4559 X-7 & 164.60 & 6245 \\
4 & 5256 & 3XMM J133000.9+471343 & MESSIER 051 & M51 ULX-7 & 493.53 & 6339 \\
5 & 28355 & 3XMM J151558.6+561810 & NGC 5907 & NGC 5907 ULX-1 & 779.24 & 14752 \\
6 & 132071 & 3XMM J005504.8-374143 & NGC 0300 & NGC 300 ULX-1 & 141.07 & 18916 \\
7 & 1840 & 4XMM J235751.0-323726 & NGC7793 & NGC 7793 P13 & 412.84 & 23022 \\
8$^{\rm b}$ & - & - & - & Circ ULX-5 & 301.67 & 23665 \\
9$^{\rm b}$ & - & - & - & NGC 1313 X-2 & 883.13 & 28853 \\
10$^{\rm b}$ & - & - & - & Holmberg IX X-1 & 156.46 & 30822 \\
11$^{\rm b}$ & - & - & - & M33 X-8 & 264.53 & 53962 \\
12 & 894 & 3XMM J031819.9-662910 & NGC 1313 & NGC 1313 X-1 & 884.82 & 59729 \\
\hline
\\
\multicolumn{7}{l}{\parbox{0.9\linewidth}{Notes: Sources with IAU names starting with 3XMM are from to the \cite{Earnshaw_ULX_catalogue} catalogue, while those starting with 4XMM are from the \cite{walton_2022} catalogue.}} \\
\multicolumn{7}{l}{\parbox{0.9\linewidth}{$^{\rm a}$ This column lists the SrcID assigned in the catalogue of \cite{Earnshaw_ULX_catalogue} or the record number assigned in the catalogue of \cite{walton_2022}, which does not assign a specific SrcID to each source detection.}} \\
\multicolumn{7}{l}{\parbox{0.9\linewidth}{$^{\rm b}$ Sources not listed in either catalogue but included as well-known ULX candidates from the literature.}} \\
\end{tabular}
\end{table}

\begin{table*} 
\centering
\caption{Detailed list of XMM-Newton observations used in this work for the selected bright, well-known ULX Sources and ULXPs in the search for the iron emission line.}
\label{tab:observation_log}
\resizebox{0.96\columnwidth}{!}{
\fontsize{6}{7}\selectfont
\begin{tabularx}{\textwidth}{X X}
\begin{tabular}{|ccccc|}
\hline
\textbf{Source} & \textbf{ObsID} & \textbf{MJD} & \textbf{Src Rgn} & \textbf{GTE (ks)} \\
\hline
M51 ULX-7 & 0112840201 & 52654.57 & $20^{''}$ & 17.12 \\
   & 0212480801 & 53552.29 & $20^{''}$ & 20.86 \\
   & 0303420101 & 53875.29 & $20^{''}$ & 28.85 \\
   & 0303420201 & 53879.48 & $20^{''}$ & 20.66 \\
   & 0677980701 & 55719.22 & $20^{''}$ & 3.84 \\
   & 0824450901 & 58251.92 & $20^{''}$ & 64.73 \\
   & 0830191501 & 58282.10 & $20^{''}$ & 51.59 \\
   & 0830191601 & 58284.09 & $20^{''}$ & 51.66 \\
   & 0852030101 & 58675.49 & $20^{''}$ & 58.78 \\
   & 0883550201 & 59542.39 & $20^{''}$ & 70.49 \\
   & 0883550301 & 59586.26 & $20^{''}$ & 104.95 \\
\hline 
NGC 300 ULX-1  & 0791010101 & 57739.40 & $30^{''}$ & 96.29 \\
   & 0791010301 & 57741.39 & $30^{''}$ & 44.78 \\
\hline 
NGC 1313 X-2 & 0150280301 & 52994.10 & $30^{''}$ & 7.23 \\
   & 0150280601 & 53012.16 & $30^{''}$ & 5.86 \\
   & 0205230301 & 53161.27 & $30^{''}$ & 8.69 \\
   & 0205230501 & 53332.31 & $30^{''}$ & 12.49 \\
   & 0205230601 & 53408.50 & $30^{''}$ & 8.48 \\
   & 0301860101 & 53800.71 & $30^{''}$ & 17.18 \\
   & 0405090101 & 54024.01 & $30^{''}$ & 71.01 \\
   & 0693850501 & 56277.68 & $30^{''}$ & 67.83 \\
   & 0693851201 & 56283.67 & $30^{''}$ & 54.29 \\
   & 0722650101 & 56451.24 & $30^{''}$ & 9.09 \\
   & 0742490101 & 57111.19 & $30^{''}$ & 66.46 \\
   & 0742590301 & 56843.97 & $30^{''}$ & 53.52 \\
   & 0764770101 & 57361.20 & $30^{''}$ & 32.88 \\
   & 0764770401 & 57470.95 & $30^{''}$ & 13.25 \\
   & 0782310101 & 57669.90 & $30^{''}$ & 73.97 \\
   & 0794580601 & 57841.25 & $30^{''}$ & 13.97 \\
   & 0803990101 & 57918.89 & $30^{''}$ & 106.32 \\
   & 0803990201 & 57924.91 & $30^{''}$ & 109.23 \\
   & 0803990301 & 57996.68 & $30^{''}$ & 28.89 \\
   & 0803990401 & 57998.75 & $30^{''}$ & 24.19 \\
   & 0803990501 & 58094.47 & $30^{''}$ & 53.65 \\
   & 0803990601 & 58096.46 & $30^{''}$ & 44.65 \\
\hline 
NGC 5907 ULX-1 & 0145190101 & 52698.59 & $30^{''}$ & 7.07 \\
   & 0145190201 & 52690.67 & $30^{''}$ & 10.14 \\
   & 0673920301 & 55966.59 & $30^{''}$ & 12.82 \\
   & 0729561301 & 56847.95 & $30^{''}$ & 37.57 \\
   & 0804090401 & 57939.02 & $30^{''}$ & 31.30 \\
   & 0804090501 & 57942.79 & $30^{''}$ & 32.27 \\
   & 0804090601 & 57949.26 & $30^{''}$ & 32.47 \\
   & 0804090801 & 58656.78 & $30^{''}$ & 15.40 \\
   & 0804090901 & 58658.79 & $30^{''}$ & 14.87 \\
   & 0804091001 & 58663.83 & $30^{''}$ & 35.53 \\
   & 0804091101 & 58669.41 & $30^{''}$ & 29.99 \\
   & 0804091201 & 58670.76 & $30^{''}$ & 9.62 \\
   & 0824320201 & 58646.82 & $30^{''}$ & 42.85 \\
   & 0824320301 & 58653.21 & $30^{''}$ & 40.64 \\
   & 0824320401 & 58660.79 & $30^{''}$ & 41.60 \\
   & 0824320601 & 59053.74 & $30^{''}$ & 26.45 \\
   & 0824320701 & 59159.90 & $30^{''}$ & 24.89 \\
   & 0851180701 & 58705.51 & $30^{''}$ & 48.35 \\
   & 0851180801 & 58707.45 & $30^{''}$ & 41.14 \\
   & 0884220201 & 59265.10 & $30^{''}$ & 19.87 \\
   & 0884220301 & 59271.09 & $30^{''}$ & 44.50 \\
   & 0891801501 & 59453.39 & $30^{''}$ & 49.44 \\
   & 0893810301 & 59585.18 & $30^{''}$ & 27.02 \\
   & 0913080201 & 59889.87 & $30^{''}$ & 55.30 \\
   & 0931410101 & 60152.77 & $30^{''}$ & 22.58 \\
   & 0953011001 & 60521.89 & $30^{''}$ & 25.56 \\
\hline 
NGC 7793 P13 & 0693760401 & 56621.21 & $30^{''}$ & 41.1 \\
   & 0748390901 & 57002.00 & $30^{''}$ & 41.97 \\
   & 0781800101 & 57528.58 & $30^{''}$ & 15.61 \\
   & 0804670301 & 57893.66 & $30^{''}$ & 39.89 \\
   & 0804670401 & 57904.90 & $30^{''}$ & 24.99 \\
   & 0804670601 & 57924.11 & $30^{''}$ & 21.05 \\
   & 0804670701 & 58083.00 & $30^{''}$ & 41.81 \\
   & 0823410301 & 58449.79 & $30^{''}$ & 19.35 \\
   & 0823410401 & 58479.60 & $30^{''}$ & 18.78 \\
   & 0840990101 & 58619.99 & $30^{''}$ & 10.24 \\
   & 0853981001 & 58809.36 & $30^{''}$ & 30.43 \\
   & 0861600101 & 59027.79 & $30^{''}$ & 42.42 \\
   & 0883780101 & 59363.72 & $30^{''}$ & 33.33 \\
   & 0883780201 & 59538.40 & $30^{''}$ & 13.12 \\
   & 0922840101 & 60100.91 & $30^{''}$ & 13.17 \\
   & 0922840201 & 60284.30 & $30^{''}$ & 5.58 \\
   &  & & & \\

\hline
\end{tabular}
&
\begin{tabular}{|ccccc|}
\hline
\textbf{Source} & \textbf{ObsID} & \textbf{MJD} & \textbf{Src Rgn} & \textbf{GTE (ks)} \\
\hline
Circ ULX-5 & 0111240101 & 52127.40 & $30^{''}$ & 63.96 \\
   & 0656580601 & 56717.44 & $30^{''}$ & 20.95 \\
   & 0701981001 & 56326.32 & $30^{''}$ & 28.29 \\
   & 0792382701 & 57623.73 & $30^{''}$ & 13.34 \\
   & 0824450301 & 58377.73 & $30^{''}$ & 86.30 \\
   & 0932990101 & 60342.30 & $30^{''}$ & 88.83 \\
\hline 
Holmberg II X-1 & 0112520601 & 52374.60 & $30^{''}$ & 4.64 \\
   & 0112520901 & 52535.11 & $30^{''}$ & 3.69 \\
   & 0200470101 & 53110.86 & $30^{''}$ & 32.73 \\
   & 0561580401 & 55281.41 & $30^{''}$ & 21.63 \\
   & 0724810101 & 56544.29 & $30^{''}$ & 4.65 \\
   & 0724810301 & 56552.27 & $30^{''}$ & 5.81 \\
   & 0864550201 & 59281.06 & $30^{''}$ & 18.97 \\
   & 0864550301 & 59285.05 & $30^{''}$ & 19.58 \\
   & 0864550401 & 59289.07 & $30^{''}$ & 17.60 \\
   & 0864550501 & 59293.02 & $30^{''}$ & 11.30 \\
   & 0864550601 & 59297.01 & $30^{''}$ & 17.77 \\
   & 0864550701 & 59301.00 & $30^{''}$ & 14.05 \\
   & 0864550801 & 59308.98 & $30^{''}$ & 19.55 \\
   & 0864550901 & 59313.00 & $30^{''}$ & 17.66 \\
   & 0864551101 & 59304.99 & $30^{''}$ & 10.02 \\
   & 0864551201 & 59316.96 & $30^{''}$ & 12.99 \\
\hline 
Holmberg IX X-1 & 0112521001 & 52374.74 & $30^{''}$ & 7.05 \\
   & 0112521101 & 52380.73 & $30^{''}$ & 7.64 \\
   & 0200980101 & 53274.31 & $30^{''}$ & 48.21 \\
   & 0657802201 & 55888.05 & $30^{''}$ & 12.17 \\
   & 0693850801 & 56223.20 & $30^{''}$ & 5.69 \\
   & 0693851001 & 56227.19 & $30^{''}$ & 3.73 \\
   & 0693851101 & 56247.14 & $30^{''}$ & 2.67 \\
   & 0693851701 & 56243.15 & $30^{''}$ & 7.14 \\
   & 0693851801 & 56245.14 & $30^{''}$ & 6.59 \\
   & 0870900201 & 59169.40 & $30^{''}$ & 17.87 \\
   & 0870930101 & 59139.51 & $30^{''}$ & 15.50 \\
   & 0870930301 & 59151.44 & $30^{''}$ & 7.70 \\
   & 0870930501 & 59171.38 & $30^{''}$ & 14.50 \\
\hline 
NGC 1313 X-1 & 0150280301 & 52994.10 & $30^{''}$ & 7.22 \\
   & 0150280601 & 53012.16 & $30^{''}$ & 5.85 \\
   & 0205230301 & 53161.27 & $30^{''}$ & 8.69 \\
   & 0205230501 & 53332.31 & $30^{''}$ & 12.48 \\
   & 0205230601 & 53408.50 & $30^{''}$ & 8.33 \\
   & 0301860101 & 53800.71 & $30^{''}$ & 17.12 \\
   & 0405090101 & 54024.01 & $30^{''}$ & 71.63 \\
   & 0693850501 & 56277.68 & $30^{''}$ & 67.84 \\
   & 0693851201 & 56283.67 & $30^{''}$ & 54.31 \\
   & 0722650101 & 56451.24 & $30^{''}$ & 9.09 \\
   & 0742490101 & 57111.19 & $30^{''}$ & 66.40 \\
   & 0742590301 & 56843.97 & $30^{''}$ & 53.68 \\
   & 0764770101 & 57361.20 & $30^{''}$ & 32.74 \\
   & 0764770401 & 57470.95 & $30^{''}$ & 13.15 \\
   & 0782310101 & 57669.90 & $30^{''}$ & 73.99 \\
   & 0794580601 & 57841.25 & $30^{''}$ & 14.01 \\
   & 0803990101 & 57918.89 & $30^{''}$ & 106.81 \\
   & 0803990201 & 57924.91 & $30^{''}$ & 109.70 \\
   & 0803990301 & 57996.68 & $30^{''}$ & 29.13 \\
   & 0803990401 & 57998.75 & $30^{''}$ & 24.41 \\
   & 0803990501 & 58094.47 & $30^{''}$ & 53.58 \\
   & 0803990601 & 58096.46 & $30^{''}$ & 44.66 \\
\hline 
M33 X-8 & 0102640601 & 52095.66 & $40^{''}$ & 2.31 \\
   & 0102642001 & 52136.35 & $40^{''}$ & 8.02 \\
   & 0102642101 & 52299.50 & $40^{''}$ & 8.97 \\
   & 0102642301 & 52301.43 & $40^{''}$ & 8.98 \\
   & 0141980301 & 52845.33 & $40^{''}$ & 5.68 \\
   & 0141980601 & 52662.84 & $40^{''}$ & 10.61 \\
   & 0141980801 & 52682.65 & $40^{''}$ & 5.57 \\
   & 0650510101 & 55386.31 & $40^{''}$ & 68.08 \\
   & 0650510201 & 55388.31 & $40^{''}$ & 65.46 \\
   & 0800350101 & 57955.35 & $40^{''}$ & 17.40 \\
   & 0800350201 & 57957.79 & $40^{''}$ & 18.34 \\
   & 0800350301 & 57967.32 & $40^{''}$ & 18.98 \\
   & 0831590201 & 58681.68 & $40^{''}$ & 10.78 \\
   & 0831590401 & 58677.73 & $40^{''}$ & 15.35 \\
\hline 
NGC 4559 ULX-7 & 0152170501 & 52786.13 & $30^{''}$ & 32.33 \\
   & 0842340201 & 58650.8 & $30^{''}$ & 56.42 \\
   & 0883960201 & 59743.86 & $30^{''}$ & 75.85 \\    
\hline
IC 342 X-2 & 093640901  & 51951.06 & $20^{''}$ & 4.80 \\    
   & 0206890201 & 53234.83 & $30^{''}$ & 16.26 \\    
   & 0693850601 & 56150.85 & $30^{''}$ & 32.77 \\
   & 0693851301 & 56156.84 & $30^{''}$ & 31.56\\
\hline 
\end{tabular}
\end{tabularx}
}
\end{table*}

\clearpage
\onecolumngrid
\renewcommand{\thetable}{B\arabic{table}} 
\renewcommand{\theHtable}{B\arabic{table}}
\setcounter{table}{0}                     

\renewcommand{\thefigure}{B\arabic{figure}} 
\renewcommand{\theHfigure}{B\arabic{figure}} 
\setcounter{figure}{0}                     
\section{Supplementary Information}
\begin{figure*}[!h]
    \centering
    \includegraphics[height=4.8cm, width=0.32\linewidth]{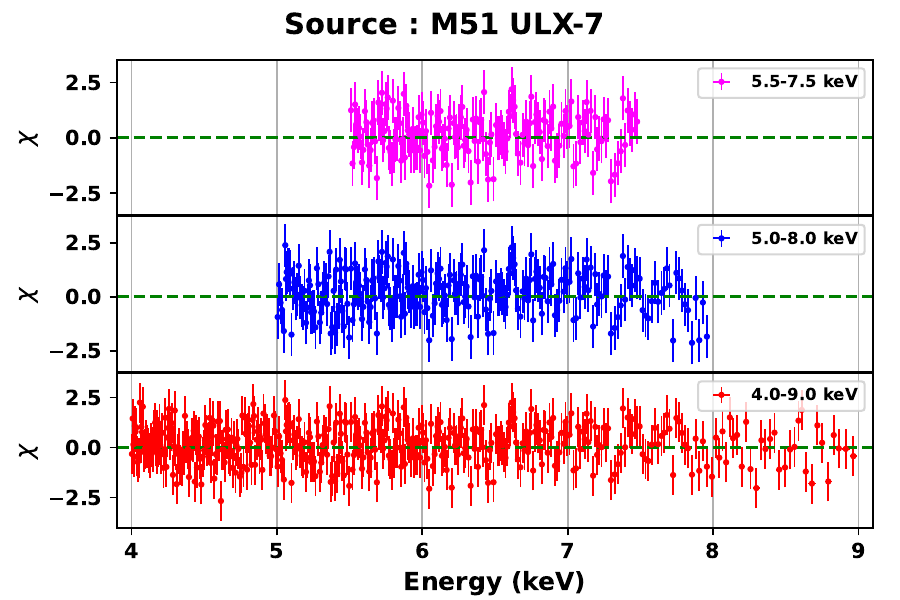}
    \includegraphics[height=4.8cm, width=0.32\linewidth]{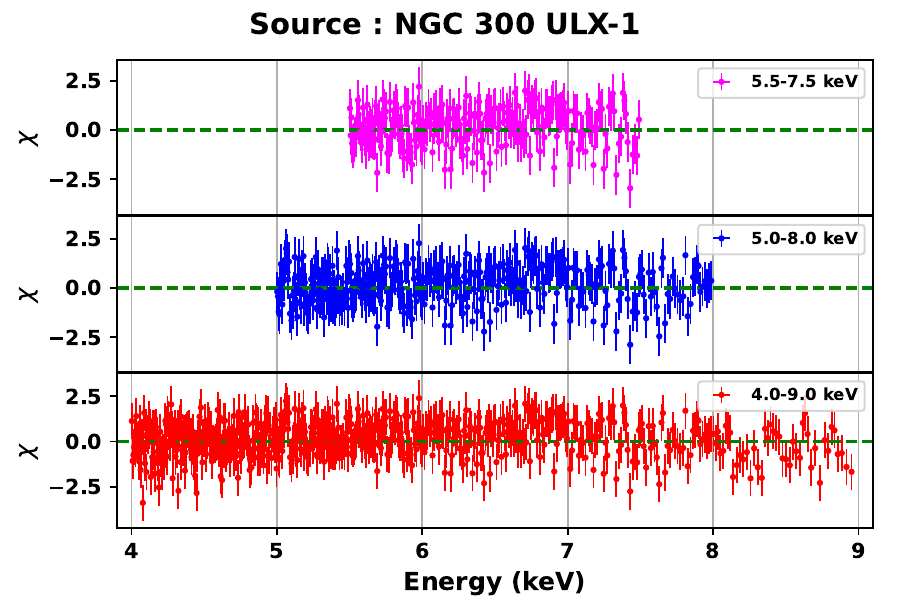}
    \vspace{0.1cm}
    \includegraphics[height=4.8cm, width=0.32\linewidth]{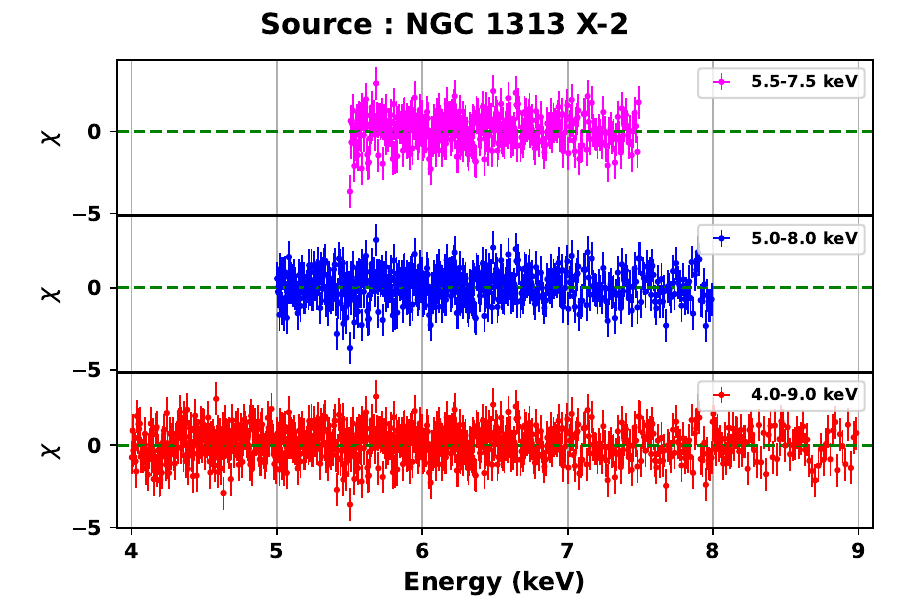}
    \includegraphics[height=4.8cm, width=0.32\linewidth]{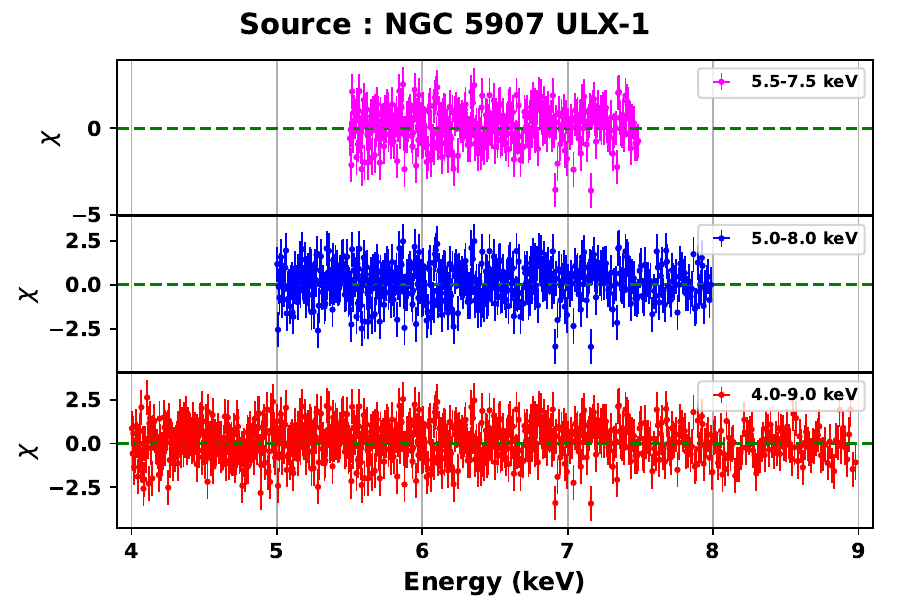}
    \vspace{0.1cm}
    \includegraphics[height=4.8cm, width=0.32\linewidth]{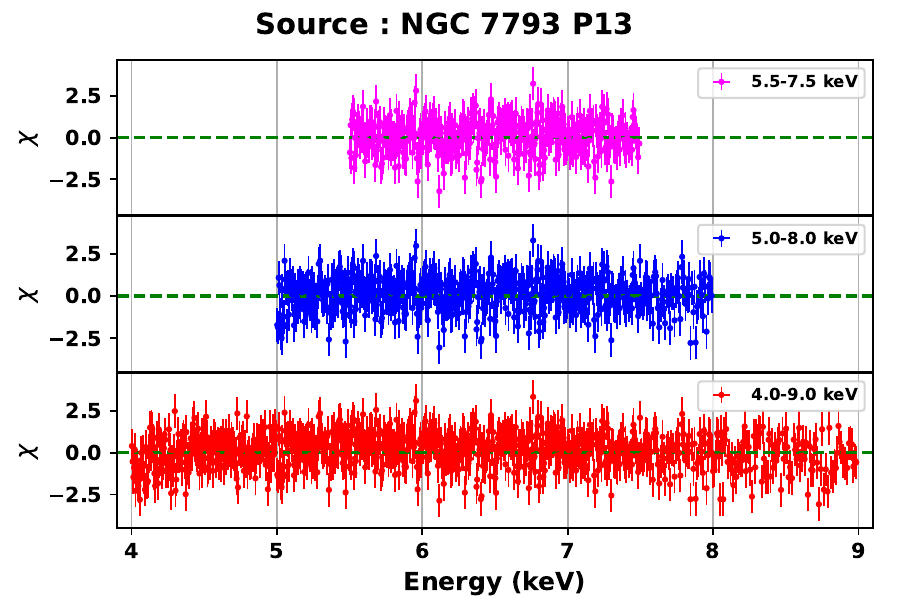}
    \includegraphics[height=4.8cm, width=0.32\linewidth]{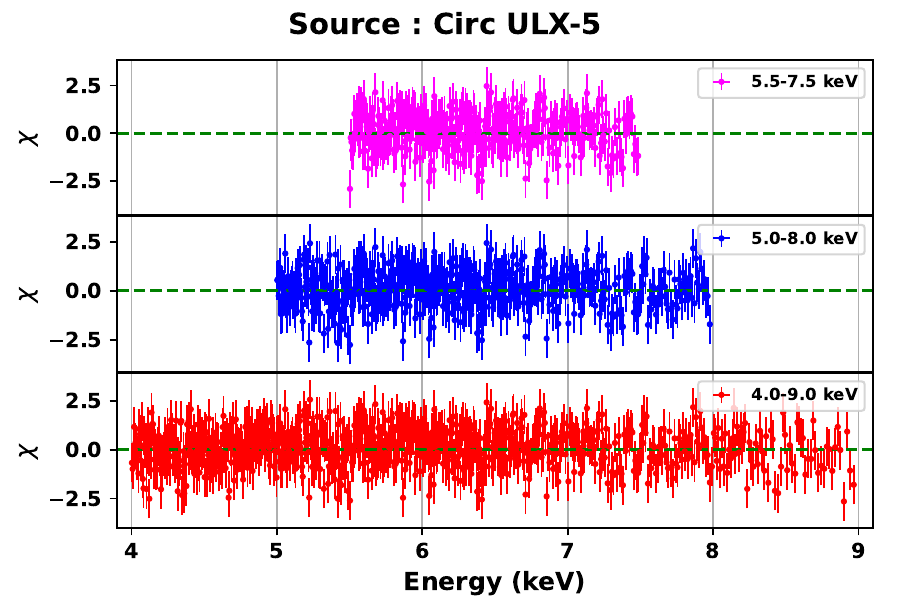}
    \vspace{0.1cm}
    \includegraphics[height=4.8cm, width=0.32\linewidth]{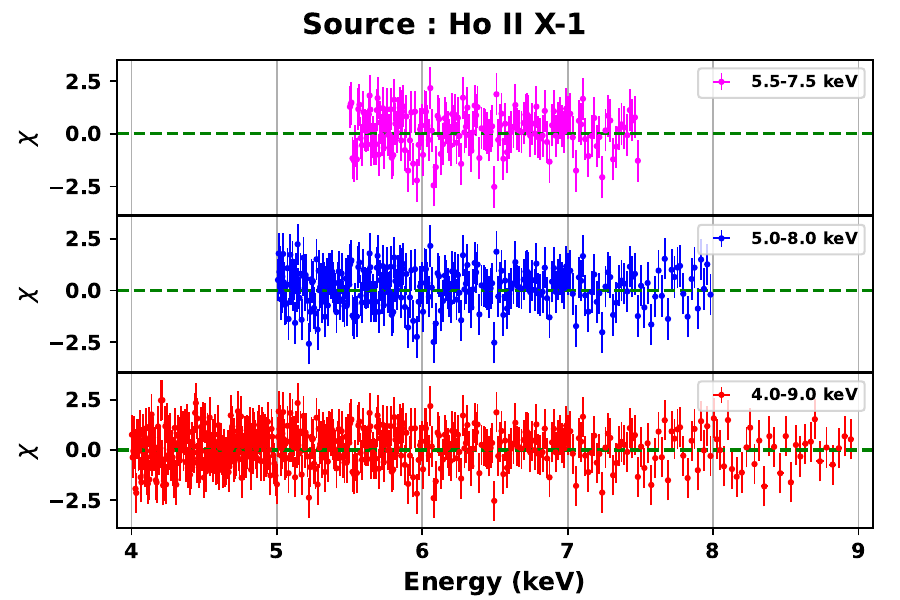}
    \includegraphics[height=4.8cm, width=0.32\linewidth]{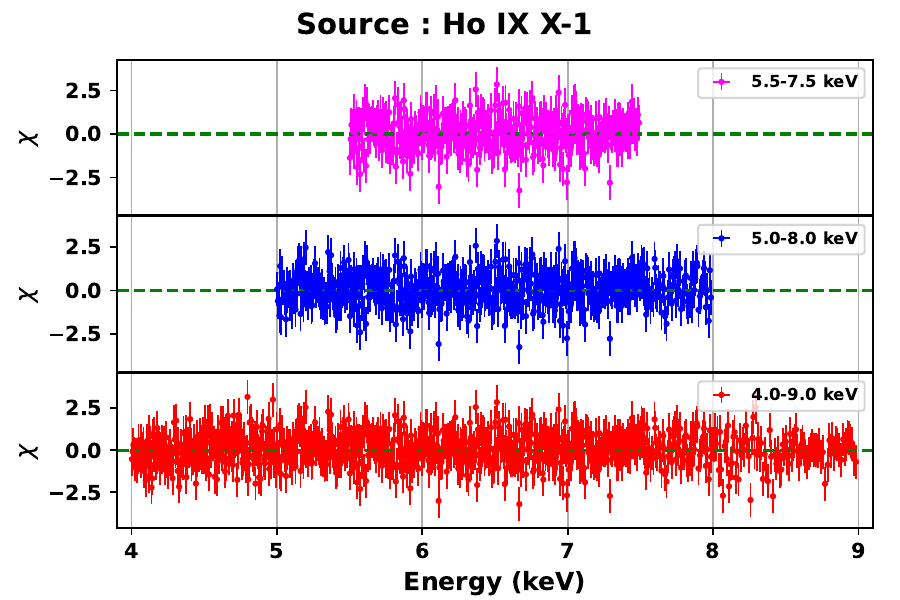}
    \vspace{0.1cm}
    \includegraphics[height=4.8cm, width=0.32\linewidth]{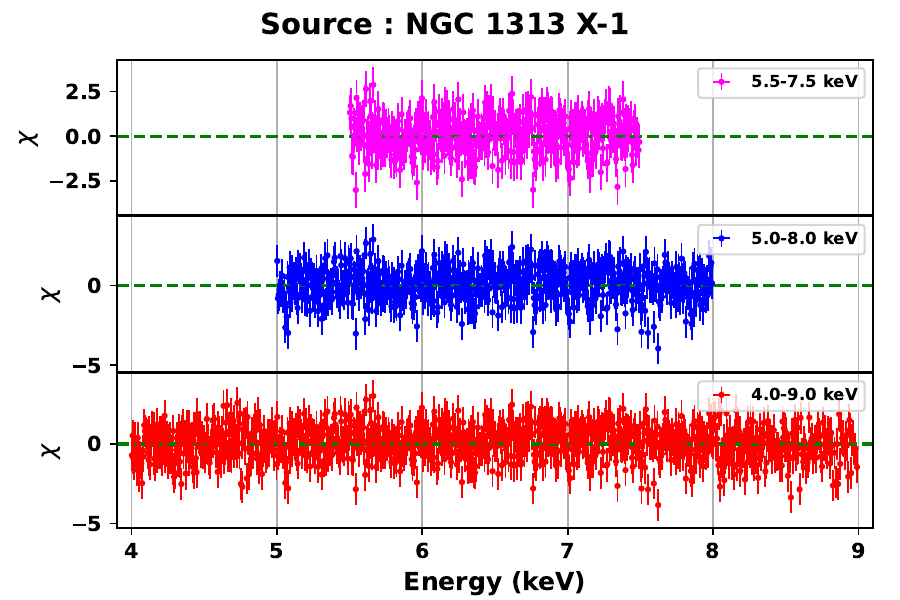}
    \includegraphics[height=4.8cm, width=0.32\linewidth]{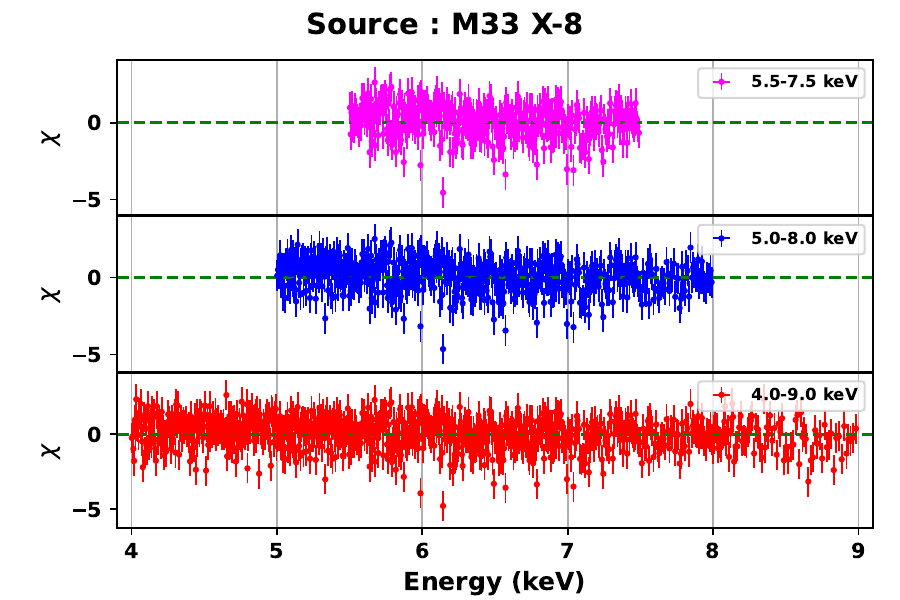}
    \vspace{0.1cm}
    \includegraphics[height=4.8cm, width=0.32\linewidth]{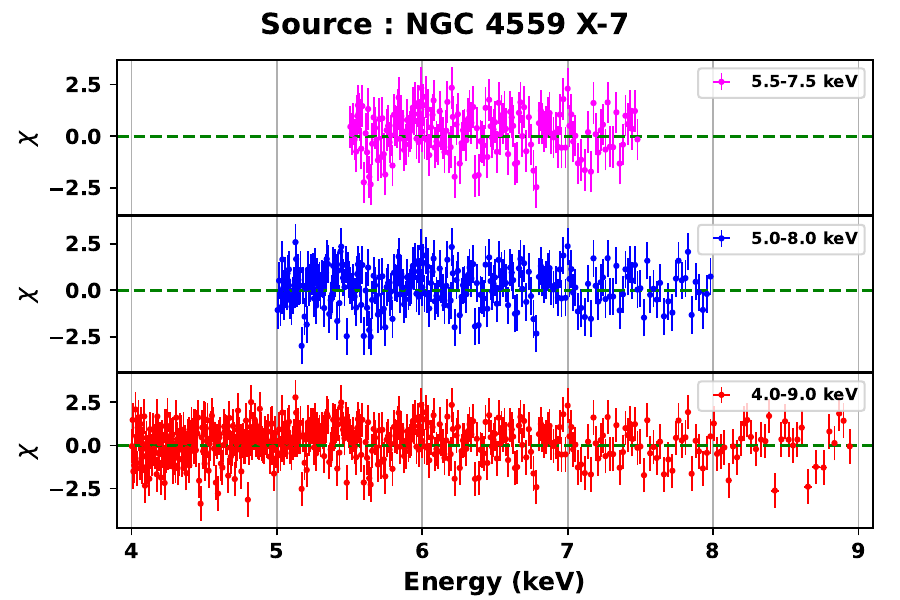}
    \includegraphics[height=4.8cm, width=0.32\linewidth]{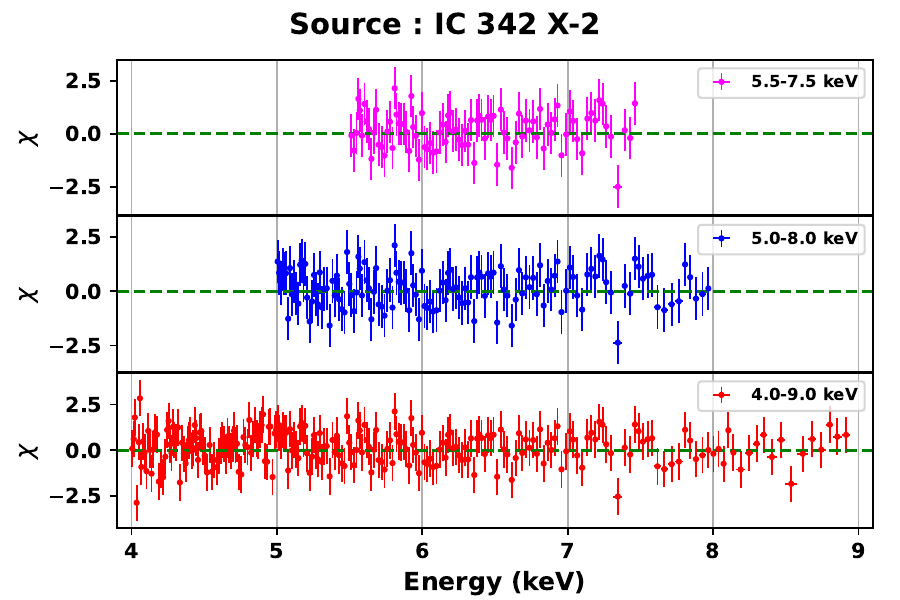}

    \caption{Residuals from the best-fit model for the combined spectra of individual sources, shown over three energy ranges: 5.5--7.5 keV, 5--8 keV, and 4--9 keV. For M51 ULX-7 only, a high-energy cutoff power-law model is used for the fit over the 4--9 keV energy range.}
    \label{fig:spec}
    
\end{figure*}

\begin{figure*}
    \centering

    \centering
    \includegraphics[width=0.48\linewidth]{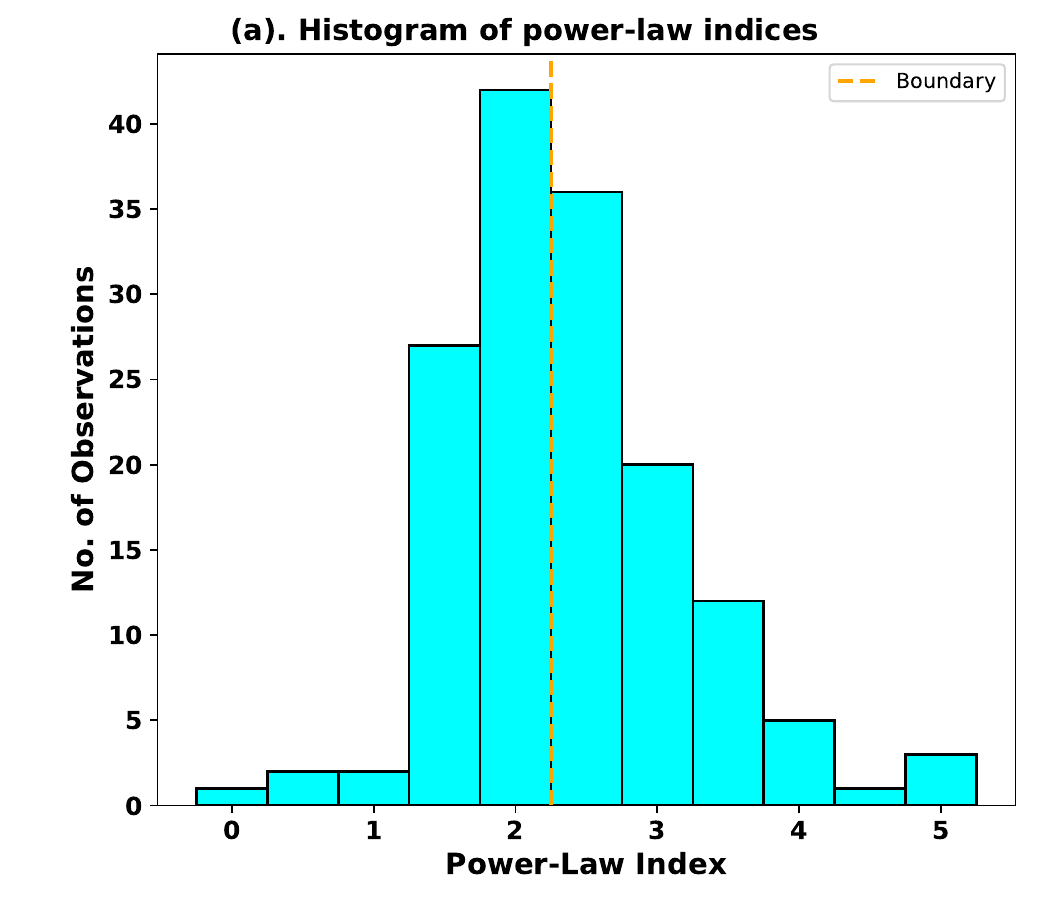}
    \vspace{0.3cm}
    \includegraphics[width=0.48\linewidth]{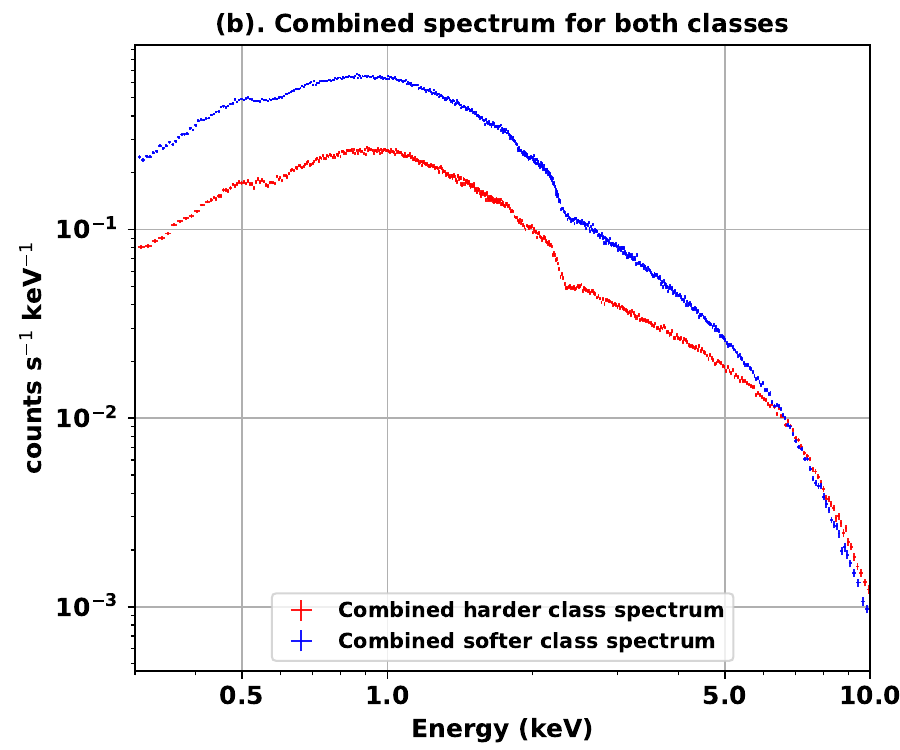}
    \includegraphics[width=0.48\linewidth]{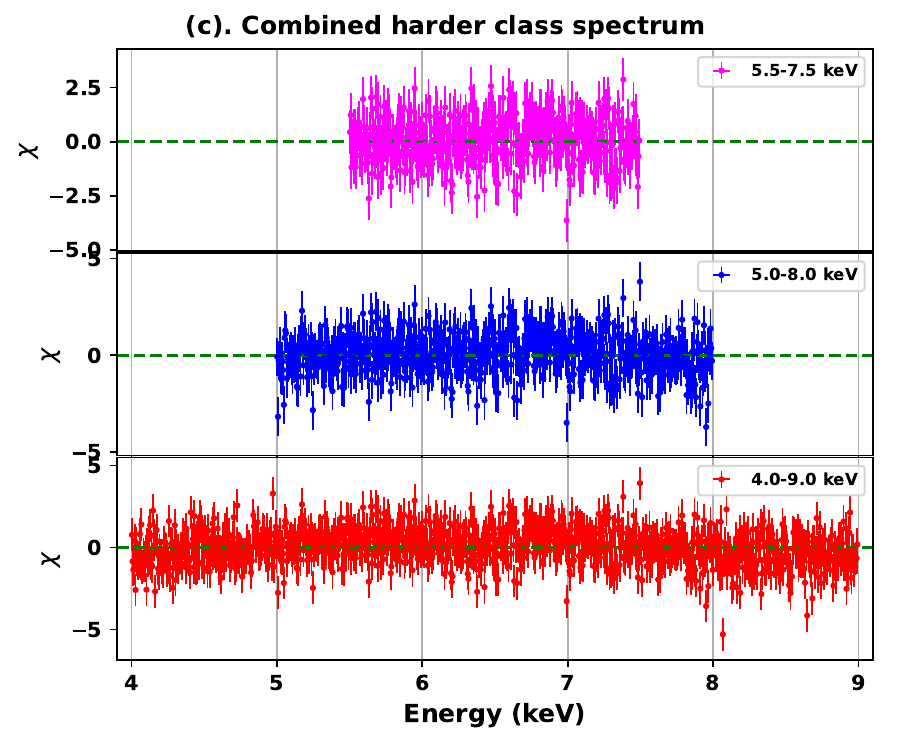}
    \includegraphics[width=0.48\linewidth]{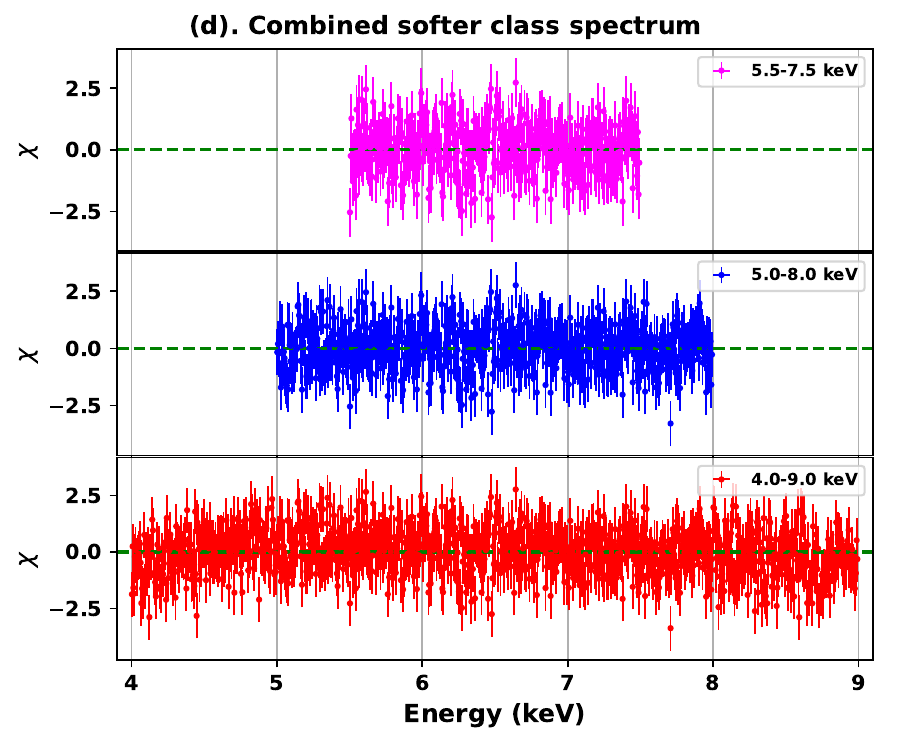}
    \vspace{0.1cm}

    \caption{(a) Histogram of the photon indices for the individual observations, used to determine the boundary between the harder and softer spectral classes for subsequent combination. (b) Comparison of the combined harder and softer class spectra. (c) and (d) Residuals from the best-fit models for the harder and softer class spectra, respectively, after fitting the spectra over the specified energy ranges.}
    \label{fig:all_4_less_greater_than}
\end{figure*}

The supplementary information presented here complements the analyses described in Sections~\ref{search_Fe_line} and \ref{robustness_Fe_line}. Fig. \ref{fig:spec} presents the residuals from the iron-line search for individual ULXs, while Fig.~\ref{fig:all_4_less_greater_than} shows the distribution of photon indices used to define the harder and softer spectral classes in the robustness analysis.


\end{document}